\documentclass[10pt,journal,compsoc]{IEEEtran}

\ifCLASSOPTIONcompsoc
  \usepackage[nocompress]{cite}
\else
  \usepackage{cite}
\fi
\ifCLASSINFOpdf
\else
\fi

\usepackage{algpseudocode}
\usepackage{amsmath,amsfonts}
\usepackage{amsmath,amssymb,amsfonts}

\usepackage{array}

\usepackage{amsmath,amsfonts}
\usepackage{amsmath,amssymb,amsfonts}
\usepackage{algpseudocode}
\usepackage{algorithm}
\usepackage{array}
\usepackage{graphicx}
\usepackage{caption}
\usepackage{multirow}

\usepackage{url}

\usepackage{fancyhdr}
\usepackage{kantlipsum}
\usepackage{eso-pic}

\begin{document}
	
\AddToShipoutPictureBG*{%
	\AtPageLowerLeft{%
		\setlength\unitlength{1in}%
		\hspace*{\dimexpr0.5\paperwidth\relax}
		\makebox(0,0.63)[c]{This work has been submitted to the IEEE for possible publication.}
		\makebox(0,0.3)[c]{Copyright may be transferred without notice, after which this version may no longer be accessible.}
}}
	
%
\title{Enhancing Industrial Cybersecurity: SoftHSM Implementation on SBCs for Mitigating MITM Attacks}
%
%
%
%

\author{Joshua~Tito~Amael,~\IEEEmembership{Member,~IEEE,}
        Jazi~Eko~Istiyanto,~\IEEEmembership{Member,~IEEE,}
    and~Oskar~Natan,~\IEEEmembership{Member,~IEEE}
\IEEEcompsocitemizethanks{\IEEEcompsocthanksitem Joshua Tito Amael is with the Department of Computer Science and Electronics, Universitas Gadjah Mada, Yogyakarta 55281, Indonesia (e-mail: joshua.tito.amael@mail.ugm.ac.id). 
\IEEEcompsocthanksitem Jazi Eko Istiyanto is with the Department of Computer Science and Electronics, Universitas Gadjah Mada, Yogyakarta 55281, Indonesia (e-mail: jazi@ugm.ac.id).
\IEEEcompsocthanksitem Oskar Natan is with the Department of Computer Science and Electronics, Universitas Gadjah Mada, Yogyakarta 55281, Indonesia (e-mail: oskarnatan@ugm.ac.id).}
\thanks{Manuscript received xx xx, xxxx; revised xx xx, xxxx.\\
Corresponding Author: Jazi Eko Istiyanto}}

%
%

\markboth{IEEE Transactions on Dependable and Secure Computing,~Vol.~xx, No.~xx, xxxx~xxxx}%
{Shell \MakeLowercase{\textit{et al.}}: Bare Advanced Demo of IEEEtran.cls for IEEE Computer Society Journals}
%



\IEEEtitleabstractindextext{%
\begin{abstract}
The rapid growth of industrial technology, driven by automation, IoT, and cloud computing, has also increased the risk of cyberattacks, such as Man-in-the-Middle (MITM) attacks. A standard solution to protect data is using a Hardware Security Module (HSM), but its high implementation cost has led to the development of a more affordable alternative: SoftHSM. This software-based module manages encryption and decryption keys using cryptographic algorithms. This study simulates the use of SoftHSM on a single-board computer (SBC) to enhance industrial system security and cost-effectively mitigate MITM attacks. The security system integrates AES and RSA cryptographic algorithms, with SoftHSM handling RSA key storage. The results show that HSM protects RSA private keys from extraction attempts, ensuring data security. In terms of performance, the system achieved an average encryption time of 3.29 seconds, a slot access time of 0.018 seconds, and a decryption time of 2.558 seconds. It also demonstrated efficient memory usage, with 37.24\% for encryption and 24.24\% for decryption, while consuming 5.20 V and 0.72 A during processing.
\end{abstract}

\begin{IEEEkeywords}
Cryptographic, SoftHSM, Cyber Security, Authentication, MITM
\end{IEEEkeywords}}

\maketitle

\IEEEdisplaynontitleabstractindextext

%
\IEEEpeerreviewmaketitle

\ifCLASSOPTIONcompsoc
\IEEEraisesectionheading{\section{Introduction}\label{sec:introduction}}
\else
\section{Introduction}
\label{sec:introduction}
\fi

\IEEEPARstart{T}{h}e development of industrial automation today is massive, with many industries being supported by various technologies such as robotics, 3D design, IoT, Big Data, and data analytics \cite{frank_industry_2019} \cite{dalenogare_expected_2018} \cite{regla_performance_2022}. Industries today tend to rely more on high automation concepts, utilizing automated machines with minimal use of workers \cite{dinlersoz_automation_2023} \cite{yuan_leakage_2024}. This condition is supported by the fact that using machines makes work more efficient in decision-making and adapting without the intervention of workers \cite{schuh_industrie_nodate}. However, the advancement of automation technology in the industry also has negative impacts, including increased cyberattacks on industries. In 2021, there were 1.6 billion anomalies recorded in industrial network traffic. Additionally, it was noted that the manufacturing sector experienced the most attacks, reaching 23.2\%. As of September 2022, there were 1,116,402 anomalies in network traffic on industrial electronic systems \cite{amael_securing_2024}. The growing number of cyberattacks has indirectly increased the number and variety of such attacks \cite{r1} \cite{wang_file_2011}. 

\begin{figure} [h] 
    \centering
    \includegraphics[width = 8.7 cm]{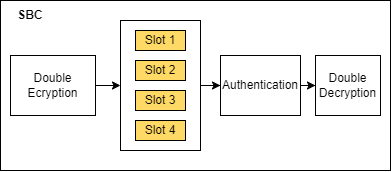}
 \caption{System Overview}
    \label{fig/overview}
\end{figure}

The main goal of a cyberattack is to obtain individual information, such as critical data within an industry, or even to damage the industry’s systems \cite{r1}. One of the prevalent cyberattacks is the man-in-the-middle (MITM) attack. MITM is a computer-based attack where a third party impersonates one of the parties in a two-way communication, tricking one party into believing they are communicating with the other \cite{mallik_man---middle-attack_2019}. In this scenario, the attacker can listen to the communication between the two unsuspecting parties to collect information \cite{saed_detection_2022}. MITM becomes even more dangerous when the attacker inserts malware or a hardware trojan that can harm industrial machines.

The Hardware Security Module (HSM) has emerged as an effective solution to mitigate these challenges. An HSM is a specialized hardware device designed to perform cryptographic operations, including the generation and secure storage of private-public key pairs and other sensitive cryptographic data \cite{hupp_module-ot_2020} \cite{bathalapalli_itpm_2023} \cite{benhani_security_2019}. It excels in key management, encryption, and decryption tasks while safeguarding this information from unauthorized access, making data extraction highly challenging \cite{mulder_trends_2023}. Despite its robust security benefits, implementing HSMs involves significant costs and requires specialized hardware and integration efforts. However, the investment is often justified by the enhanced security and risk mitigation it provides, especially in industries where data protection is critical \cite{balan_puf-based_2020}.

In this research, a software system named SoftHSM was developed for use on a SBC. SoftHSM is designed to manage cryptographic keys and is integrated with a double encryption algorithm for enhanced security. The system operates on a key management framework, securely storing encrypted keys in dedicated key slots within SoftHSM.

The novelties of our research include:

\begin{itemize} 
\item This research realizes a low-cost HSM with a high level of key security, providing an alternative for HSM development.
\item The research incorporates four security layers: double encryption, SoftHSM key slots, and authentication.
\end{itemize}

An overview of the entire system is shown in Fig. \ref{fig/overview}. The rest of this paper is organized as follows.  In Section II, we do a comprehensive review of some related works that also inspire this research. In Section III, we explain our proposed model, especially SoftHSM (hybrid encryption and Slot System) and system evaluation. Then, we analyze the result to understand Sistem's time consumption, memory consumption, power consumption, data integrity and attacking simulation in Section IV. Finally, we conclude our findings in Section V, along with some suggestions for future studies.

\section{Related Works} \label{relatedworks}
This section reviews related works on cryptography, hardware security modules, key management systems, and Authentication. Then, we point out the key ideas that inspire our work and serve as an objective for comparative study.

\subsection{Cyptographic}
Cryptography is an essential aspect of this research, serving as one of the key data security processes. This study is inspired by the research of Jaspin et al., which introduced a data protection method in the cloud by integrating AES and RSA algorithms to create a hybrid encryption system. This combination is based on initial testing, which demonstrated that AES and RSA offer the most robust security, fastest execution times, and superior data integrity while also generating ciphertext that matches the length of the plaintext, outperforming other encryption algorithms \cite{jaspin_efficient_2021}. This research has inspired us to combine AES and RSA algorithms due to their efficiency and firm performance.

Prawira et al. conducted a study on securing SMS messages using an Arduino microcontroller, incorporating an AES cryptography algorithm enhanced with a chaotic logistic-based PRBG algorithm for key generation. The PRBG algorithm introduces a random bit sequence as input to generate random bits for AES encryption. Additionally, the system employs the SHA algorithm to ensure data integrity through preprocessing, padding, and round computations to produce a consistent hash value. Performance tests revealed that sending encrypted messages took an average of 2.118 seconds over ten trials, whereas unencrypted messages averaged a transmission time of 1.539 seconds \cite{prawira_p_secure_2020}.

Yilmaz and Ozdemir evaluated IoT devices using cryptographic algorithms, such as AES, Blowfish, DES, SHA, ECC, and RSA. Their tests assessed the energy consumption performance of the IoT devices by measuring both MiB/Second and Operation/Second, which provided insight into the number of mebibytes processed by the devices within one-second \cite{yilmaz_performance_2018}.

Homma et al. proposed a formal approach for designing cryptographic processor datapaths using arithmetic circuits over Galois fields (GF). This approach models GF arithmetic circuits as hierarchical graph structures, where the nodes represent sub-circuits with functions defined by arithmetic formulas over GFs, and the edges indicate data dependencies between nodes. This structure enables formal verification via symbolic computation techniques, such as polynomial reduction and Gröbner bases. The method's effectiveness is illustrated through the experimental design of a 128-bit AES datapath, which includes multiplicative inversion circuits over composite fields \cite{homma_toward_2014}.

Building on insights from previous research, this study aims to advance the development of a hybrid cryptographic encryption system that enhances both security and performance for related applications. This research implements a cryptographic solution on an SBC and integrates it with SoftHSM. The choice of AES and RSA algorithms is grounded in earlier findings, which showed that this combination offers a superior balance of robust security, high-speed data processing, and strong data integrity. By harnessing these strengths, the study seeks to further optimize the protection and efficiency of cryptographic systems for real-world applications.

\subsection{Hardware Security Module}\label{AA}
Mulder et al. describe a Hardware Security Module (HSM) as a specialized device used for cryptographic operations, particularly for generating and storing private-public key pairs and related secret information. HSMs can securely manage encryption keys, perform encryption and decryption, and store sensitive data, making extraction challenging. Additionally, some HSMs come with physical protections like tamper-proof features, which turn off the device if tampering is detected \cite{mulder_trends_2023-1}.

Rady et al. developed an HSM based on memristors for securing IoT devices. Memristors were selected for their unique, unpredictable characteristics, functioning as Physical Unclonable Functions (PUF), which cannot be replicated. Each memristor’s distinct behavior generates an AES key for IoT security. However, this research is limited to AES 128 keys, and the generated keys are only 3 bits in size \cite{rady_memristor-based_2019}.

Jafarzadeh studied FPGA-based HSMs that were tested against malware in the form of Hardware Trojans. The tests involved comparing power consumption, delays, and design integrity before and after the Trojan attack. The results indicated that the Trojan had little effect on power and delays but did alter the FPGA design \cite{jafarzadeh_real_2020}.

Jingwei Hu et al. explored the performance of rank-code-based cryptographic schemes on FPGA platforms, focusing on the quantum-safe Key Encapsulation Mechanism (KEM) scheme, ROLLO, which uses LRPC codes. ROLLO was a candidate in the second round of the NIST post-quantum cryptography standardization process. The implementation covers encapsulation and decapsulation operations, with some modifications from the original design. It is fully parameterized, allowing flexibility in security-level configurations using code-generation scripts \cite{hu_engineering_2023}.

Pott et al. explored data protection innovations for the autonomous vehicle industry, proposing a firmware security module (FSM) as a potential alternative to traditional Hardware Security Modules (HSMs). The FSM, built using the Infineon Aurix TC399XP microcontroller, integrates cryptographic algorithms like AES 128, SHA 256, and ECC 256 into its architecture. While the study found that FSM processing times were 1.6 to 4.5 times slower than HSM, it highlighted FSM’s greater flexibility, suggesting that software-based security solutions like FSM can serve as a viable replacement for HSMs \cite{pott_firmware_2021}. This research inspired us to develop a more affordable security system similar to an HSM, while still maintaining the same functionality.

\begin{figure*} [t] 
    \centering
    \includegraphics[width = 17.5 cm]{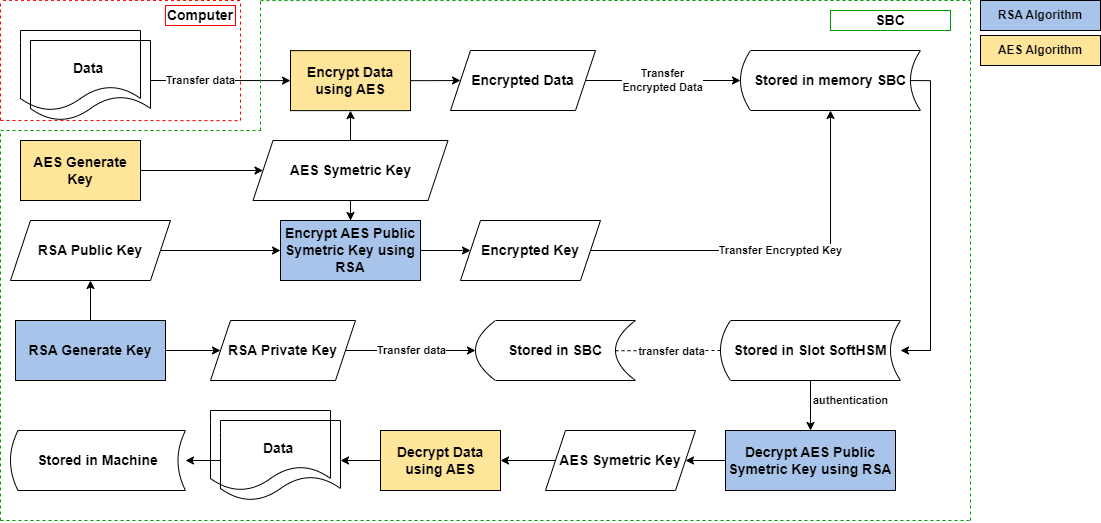}
    \caption{HSM Technical Diagram}
    \label{fig/technical diagram}
\end{figure*}

\subsection{Key Management System}
Data protection through cryptography requires a key management system (KMS) to store cryptographic keys securely. Kuzminykh et al. developed an automated KMS that integrates a server with an HSM and a database for managing, storing, and renewing cryptographic keys. This system also protects sensitive data using hardware MAX keys, preventing unauthorized server editing. Luo et al. introduced TZ-KMS, a key management system designed for cloud services on an ARM board. Though centralized, TZ-KMS improves security through user authentication and cryptographic operations while resisting DoS attacks. We developed SoftHSM, a more flexible and user-friendly tool based on these previous systems. SoftHSM functions similarly to other KMS but is explicitly tailored for hardware data security. Its open-source nature further enhances accessibility for users \cite{kuzminykh_analysis_2020} \cite{luo_tz-kms_2018} \cite{}.

\subsection{Authentication}
The authentication process in a Hardware Security Module (HSM) plays a vital role in data protection. Kim et al. developed an authentication method using Physical Unclonable Functions (PUF) combined with White Box Cryptography for secure key management in IoT devices \cite{kim_reinforcement_2022}. They also explored device authentication on the OCF Iotivity platform, using PUFs for secure cloud-based device registration. Yoon et al \cite{yoon_security_2023}. focused on mobile device security through hardware cryptographic modules, leveraging PUF for network security and authentication. Hassan et al. \cite{hassan_enhanced_2020}. tackled multiserver security threats with a dual-layered authentication system involving passwords and smart cards. Sungjin et al. examined vital drone agreement protocols but deemed them impractical for real-world use. These studies inspired us to create a data security system with an authentication scheme for IoT devices, particularly ventilators \cite{yu_cryptanalysis_2023} \cite{etibou_iot_2024} \cite{luo_fast_2024}. Building on the success of past research, we implement an authentication system in SoftHSM to secure user access to data.

\section{Method} \label{methods}
This section describes our proposed system model, Hybrid Encryption Schema, and SoftHSM, highlighting the key components that contribute to its secure operation. Then, we conduct system evaluations to ensure the system runs well in terms of load memory, time consumption, power consumption, attacking simulation, and data integrity, ensuring it meets the necessary performance benchmarks.

\subsection{Proposed Model}\label{AA}
The hardware security module system uses a single-board computer as the central processing component, connected to a PC. In this case, the Jetson Nano is the single-board computer used in the HSM. The choice of the Jetson Nano in this research is also inspired by the work of Chenxu Wang et al., who developed a trusted execution environment designed explicitly for ARM-based GPUs. Similarly, we are innovating to create a key storage system on the ARM-based Jetson Nano \cite{wang_building_2024}. The system design developed in this research is shown in Fig. \ref{fig/overview}, where the HSM will consist of four main processes: encryption, key storage or key management system, authentication, and decryption.

The design of the HSM system can be seen in Fig. \ref{fig/technical diagram}, where the PC component manually sends data to the Jetson Nano. The encryption process is first performed within the HSM using a double-layered encryption system that applies AES and RSA algorithms. In this process, the RSA private key is a crucial part that must be secured. Therefore, the data sender must create a SoftHSM slot to store the related key. This slot is created by the sender, who also sets a security PIN to ensure the slot's security. The RSA private key will then be stored in SoftHSM, like being locked in a secure compartment within the Jetson Nano. This RSA key can only be accessed when the data recipient wants to decrypt the data by entering the security PIN previously set by the sender. The process then continues with decryption, which transforms the ciphertext into plaintext.

\subsection{Hybrid Encryption Schema}
The double or hybrid encryption system combines two algorithms, AES and RSA. These algorithms are selected based on the combined capabilities, which offer fast execution, strong data integrity, and high security \cite{amael_securing_2024} \cite{cai_secfed_2024}. The data sender applies the hybrid encryption scheme to secure the data before distribution to the recipient.

Hybrid encryption begins with encrypting the data using the AES algorithm. The encryption process with AES is carried out in three stages: initial rounds, main rounds, and final rounds.

\begin{itemize}
    \item \textbf{Initial Round}\\
    This is the process of preparing the data before entering the central encryption. The initial round is performed by executing a state operation that is XORed with the first RoundKey:
  \[
  \text{State} = \text{AddRoundKey}(\text{PlainText}, \text{RoundKey})
  \]
  AddRoundKey: Each byte of the data block (plaintext) is XORed with the corresponding byte from the first round key (RoundKey), which results in:
  \[
  \text{State} = \text{PlainText} \oplus \text{RoundKey}
  \]
  The result of this process is the state that has been XORed with the round key and is ready for further processing in the main rounds, the algorithm of this system is as follows Algorithm 1:

\begin{algorithm}
\caption{Initial Round of Encryption}
\begin{algorithmic}[1]
\State \textbf{Input:} $P = \{p_1, p_2, \dots, p_n\}$ (PlainText), $K = \{k_1, k_2, \dots, k_n\}$ (RoundKey)
\State \textbf{Output:} $S = \{s_1, s_2, \dots, s_n\}$ (XORed State)
\State
\State \textbf{Step 1: AddRoundKey}
\State
\State For each byte $p_i$ of the PlainText $P$, XOR with the corresponding byte $k_i$ from the RoundKey $K$:
\[
s_i = p_i \oplus k_i \quad \text{for } i = 1, 2, \dots, n
\]
\State The XOR operation results in the state $S$:
\[
S = P \oplus K = \{p_1 \oplus k_1, p_2 \oplus k_2, \dots, p_n \oplus k_n\}
\]
\State
\State \textbf{Step 2: Update State}
\State The new state $S$ is now ready for further rounds of encryption.
\end{algorithmic}
\end{algorithm}
    
\end{itemize}

\begin{itemize}
   \item \textbf{Main Round}\\
Main Rounds are the core encryption actions applied to the data using a combination of several steps to ensure robust security. The Main Rounds consist of SubBytes, ShiftRows, MixColumns, and AddRoundKey. The following are the equations used in the Main Rounds:

First, the SubBytes transformation is applied, which substitutes each byte of the state with a corresponding value from an S-box, as shown in Equation (1):
\begin{align}
S_1 &= \text{SubBytes}(\text{State}) \tag{1}
\end{align}

Next, the ShiftRows transformation rearranges the rows of the state by shifting them cyclically, as described in Equation (2):
\begin{align}
S_2 &= \text{ShiftRows}(S_1) \tag{2}
\end{align}

Then, the MixColumns transformation mixes the columns of the state using a matrix multiplication over a Galois field, as described in Equation (3):
\begin{align}
S_3 &= \text{MixColumns}(S_2) \tag{3}
\end{align}

Finally, the AddRoundKey transformation is applied, which XORs the transformed state with the RoundKey, as shown in Equation (4):
\begin{align}
\text{State} &= \text{AddRoundKey}(S_3) \tag{4}
\end{align}

The result of these rounds is a state that becomes increasingly difficult to break, making it more secure. This algorithm is enhanced with the principle of iterative encoding. The AES algorithm's scrambling process is repeated ten times to achieve a truly random value. After completing these iterations, the process moves to the final step, the final rounds.

The algorithm of this system is as follows Algorithm 2:
\begin{algorithm}
\caption{Detailed Main Round of Encryption}
\begin{algorithmic}[1]
\State \textbf{Input:} $S = \{s_1, s_2, \dots, s_n\}$ (State), $K = \{k_1, k_2, \dots, k_n\}$ (RoundKey)
\State \textbf{Output:} Updated State $S'$
\State
\State \textbf{Step 1: SubBytes Transformation}
\State Each byte $s_i$ of $S$ is replaced by a corresponding value from an S-box. Mathematically, this is represented as:
\[
S_1 = \text{SubBytes}(S) = \left\{ S_{\text{box}}(s_1), S_{\text{box}}(s_2), \dots, S_{\text{box}}(s_n) \right\}
\]
where $S_{\text{box}}(s_i)$ represents the substitution of $s_i$ using the S-box function.
\State
\State \textbf{Step 2: ShiftRows Transformation}
\State The rows of the state matrix are shifted cyclically by a fixed offset. For a $4 \times 4$ matrix, the transformation is:
\[
S_2 = \begin{bmatrix} 
s_{00} & s_{01} & s_{02} & s_{03} \\
s_{10} & s_{11} & s_{12} & s_{13} \\
s_{20} & s_{21} & s_{22} & s_{23} \\
s_{30} & s_{31} & s_{32} & s_{33}
\end{bmatrix}
\to
\begin{bmatrix} 
s_{00} & s_{01} & s_{02} & s_{03} \\
s_{11} & s_{12} & s_{13} & s_{10} \\
s_{22} & s_{23} & s_{20} & s_{21} \\
s_{33} & s_{30} & s_{31} & s_{32}
\end{bmatrix}
\]
\State
\State \textbf{Step 3: MixColumns Transformation}
\State The state matrix is multiplied by a fixed matrix over a Galois field. For each column, the transformation is represented as:
\[
\begin{bmatrix} 
s'_{00} \\
s'_{10} \\
s'_{20} \\
s'_{30}
\end{bmatrix}
=
\begin{bmatrix} 
2 & 3 & 1 & 1 \\
1 & 2 & 3 & 1 \\
1 & 1 & 2 & 3 \\
3 & 1 & 1 & 2
\end{bmatrix}
\cdot
\begin{bmatrix} 
s_{00} \\
s_{10} \\
s_{20} \\
s_{30}
\end{bmatrix}
\]
This operation is repeated for each column of the state matrix.
\State
\State \textbf{Step 4: AddRoundKey}
\State XOR the transformed state $S_3$ with the RoundKey $K$. This operation is applied byte-wise:
\[
S' = S_3 \oplus K = \left\{ s_1' \oplus k_1, s_2' \oplus k_2, \dots, s_n' \oplus k_n \right\}
\]
where $S'$ represents the final updated state.
\State
\State \textbf{Output:} The updated state $S'$ is ready for the next round.
\end{algorithmic}
\end{algorithm}

\end{itemize}

\begin{itemize}
    \item \textbf{Final Rounds}\\
    The Final Rounds aim to produce the ciphertext, which is the final result of the encryption process. This is achieved using the following steps:
    
    First, apply the SubBytes transformation, as shown in Equation (1):
    \begin{align}
    S_1 &= \text{SubBytes}(\text{State}) \tag{1}
    \end{align}
    
    Next, apply the ShiftRows transformation, described in Equation (2):
    \begin{align}
    S_2 &= \text{ShiftRows}(S_1) \tag{2}
    \end{align}
    
    Finally, apply the AddRoundKey transformation to obtain the final state, as shown in Equation (3):
    \begin{align}
    \text{State} &= \text{AddRoundKey}(S_2, \text{RoundKey}) \tag{3}
    \end{align}
    
    The result of this process is that the state is transformed into ciphertext.

\end{itemize}

The process then continues by encrypting the AES key using the RSA algorithm. This is done by first generating the RSA private and public keys. The AES key is then encrypted in PEM (Privacy Enhanced Mail) format. The encryption is performed using the \texttt{RSA\_public\_encrypt()} function from OpenSSL, or it can be represented by the formula in Equation (4):
\begin{align}
    c &= m^e \mod n \tag{4}
\end{align}

Where:
\begin{itemize}
    \item $c$ is the ciphertext, the result of encryption, as shown in Equation (4).
    
    \item $m$ is the plaintext message that has been converted into an integer, ensuring that $0 \leq m \leq n$, as described in Equation (5):
    \begin{align}
    0 \leq m \leq n \tag{5}
    \end{align}
    
    \item $e$ is the public exponent from the public key $(e, n)$, as referenced in Equation (4).
    
    \item $n$ is the modulus, which is the product of two large prime numbers $p$ and $q$.
\end{itemize}
Thus, the value of c is the encrypted data that can be sent to the recipient. The RSA private key, which will be used for decryption, is a crucial part of RSA and will be stored in the SoftHSM Slot. The algorithm of this system is as Follow Algorithm 3:
\begin{algorithm}
\caption{Final Rounds and RSA Encryption Process}
\begin{algorithmic}[1]
\State \textbf{Input:} State $S = \{s_1, s_2, \dots, s_n\}$, RoundKey $K = \{k_1, k_2, \dots, k_n\}$
\State \textbf{Output:} Ciphertext $C$

\State
\State \textbf{Step 1: SubBytes Transformation}
\State Each byte of the state $S$ is replaced using the S-box:
\begin{align}
S_1 &= \text{SubBytes}(S) = \{ S_{\text{box}}(s_1), S_{\text{box}}(s_2), \dots, S_{\text{box}}(s_n) \} \tag{1}
\end{align}

\State \textbf{Step 2: ShiftRows Transformation}
\State The rows of the state matrix are shifted cyclically. For a 4x4 matrix, this is represented as:
\[
S_2 = \begin{bmatrix} 
s_{00} & s_{01} & s_{02} & s_{03} \\
s_{10} & s_{11} & s_{12} & s_{13} \\
s_{20} & s_{21} & s_{22} & s_{23} \\
s_{30} & s_{31} & s_{32} & s_{33}
\end{bmatrix}
\to
\begin{bmatrix} 
s_{00} & s_{01} & s_{02} & s_{03} \\
s_{11} & s_{12} & s_{13} & s_{10} \\
s_{22} & s_{23} & s_{20} & s_{21} \\
s_{33} & s_{30} & s_{31} & s_{32}
\end{bmatrix}
\tag{2}
\]

\State \textbf{Step 3: AddRoundKey Transformation}
\State XOR the transformed state $S_2$ with the RoundKey $K$, producing the final state:
\begin{align}
\text{State} &= S_2 \oplus K = \left\{ s_1' \oplus k_1, s_2' \oplus k_2, \dots, s_n' \oplus k_n \right\} \tag{3}
\end{align}

\State \textbf{Step 4: Transform State to Ciphertext}
\State The final state is transformed into ciphertext $C$:
\begin{align}
C &= \text{State} \tag{4}
\end{align}

\State \textbf{Step 5: RSA Encryption of AES Key}
\State The AES key $K$ is encrypted using the RSA algorithm. First, the AES key is converted into an integer $m$ and encrypted using the public key $(e, n)$, where:
\begin{align}
c &= m^e \mod n \tag{5}
\end{align}
\State The ciphertext $c$ is the result of RSA encryption of the AES key.

\end{algorithmic}
\end{algorithm}

\subsection{SoftHSM}
\begin{figure} [h] 
    \centering
    \includegraphics[width = 8.5 cm]{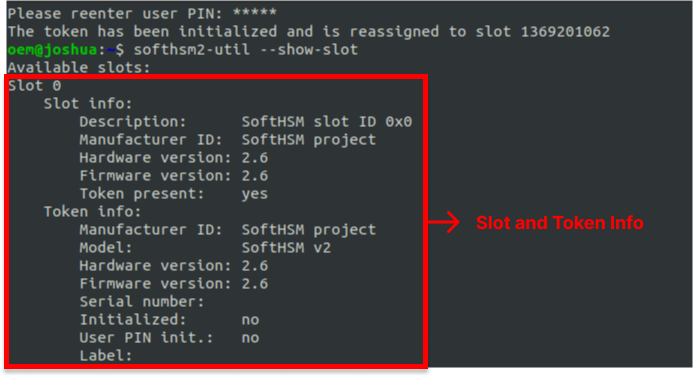}
 \caption{Interface SoftHSM}
    \label{fig/interface softhsm}
\end{figure}

SoftHSM implements cryptographic key storage that can be accessed through the PKCS11 interface. SoftHSM is developed using the PKCS11 interface, which includes cryptographic tokens, also known as cryptoki \cite{amael_securing_2024}. SoftHSM was developed to provide a cryptographic key slot storage service in this research. The interface of SoftHSM is shown in Fig. \ref{fig/interface softhsm}. Each slot within SoftHSM will contain an RSA private key, descriptive labels, manufacturer ID, hardware version, data version, model, serial number, user init PIN, and labels.

\begin{figure} [h] 
    \centering
    \includegraphics[width = 8.5 cm]{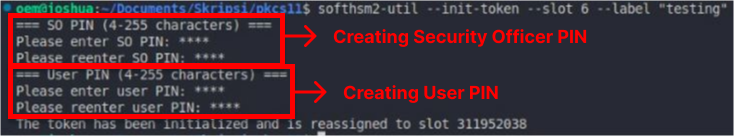}
 \caption{Slot SoftHSM Production}
    \label{fig/slot softhsm}
\end{figure}

In this research, the operation of SoftHSM begins by creating a slot within SoftHSM by entering the slot name, slot label, user PIN, and security officer PIN, as shown in Fig. \ref{fig/slot softhsm}. The user PIN can be used by the consumer to authenticate. Meanwhile, the administrator holds the security officer's PIN to perform edits on the slot and insert keys. The process of entering the PIN and creating the slot is illustrated in Fig. \ref{fig/slot softhsm}.

\begin{figure} [h] 
    \centering
    \includegraphics[width = 8.5 cm]{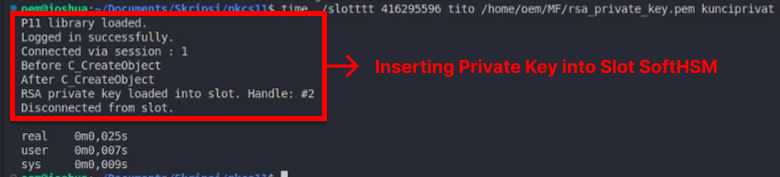}
 \caption{
The process of inserting a key into a SoftHSM slot involves}
    \label{fig/process}
\end{figure}

Next, the process continues by importing the key into the SoftHSM slot. This key import process uses the PKCS11 interface tools and the OpenSSL library. The data sender will enter the target slot number and the corresponding slot label based on the previously created slot, as shown in Fig. \ref{fig/process}.

\begin{figure} [h] 
    \centering
    \includegraphics[width = 8.5 cm]{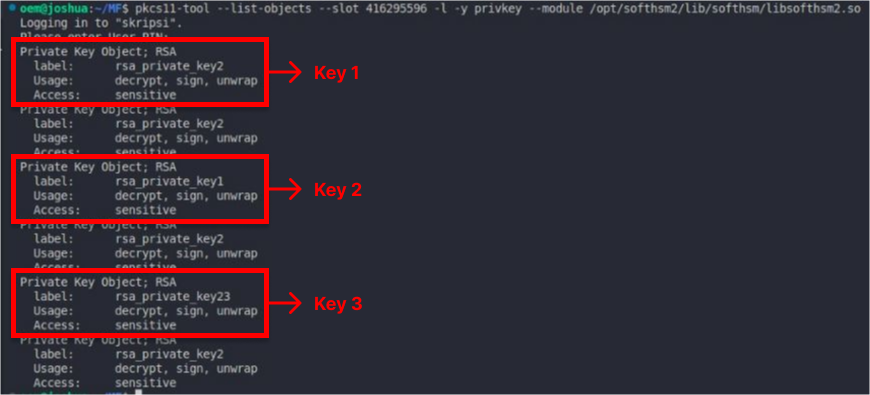}
 \caption{Key List inside of Slot SoftHSM}
    \label{fig/key list}
\end{figure}

The slot number and label will serve as unique identifiers for the slot containing the key. The private key stored within the slot will have a unique name along with its key status, as shown in Fig. \ref{fig/key list}. Additionally, the key's status will be visible, indicating whether it is protected, unprotected, or highly protected.

\begin{figure} [h] 
    \centering
    \includegraphics[width = 8.5 cm]{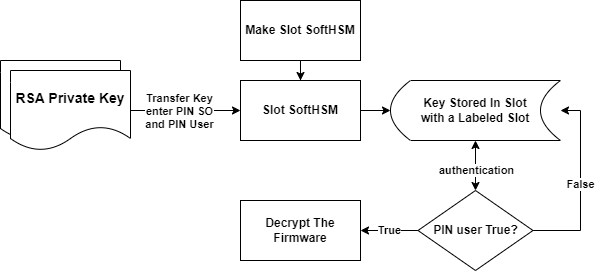}
 \caption{Slot SoftHSM Workflow}
    \label{fig/workflow}
\end{figure}

The RSA private key can only be accessed once the authentication process is completed. The data recipient carries out authentication by entering the user PIN provided by the data sender. If the entered PIN differs from the PIN stored in the SoftHSM slot, the authentication will be considered unsuccessful, and the decryption process cannot proceed. However, if the entered PIN matches the PIN stored in the slot, the authentication is booming, and the key can be used for decryption. The workflow of the SoftHSM slot is illustrated in Fig. \ref{fig/workflow}.

\subsection{Hybrid Decryption}
The final stage of this research is hybrid decryption, or the process of returning the data to its original form. The process begins with decrypting the data using the RSA algorithm to retrieve the AES key. This RSA decryption is performed using the \texttt{RSA\_private\_decrypt()} function. The decryption function contains a mathematical operation to convert the ciphertext back into plaintext as follows:

\begin{align}
m &= c^d \mod n \tag{1}
\end{align}

Where:
\begin{itemize}
    \item $c$ is the ciphertext that will be decrypted, as shown in Equation (1).
    \item $m$ is the plaintext, or the result of decryption, which will be produced as the output, also described in Equation (1).
    \item $d$ is the private exponent, known only to the message recipient (private key), and is calculated using the public key $e$ and Euler’s totient function $\phi(n)$, as shown in Equation (2):
    \begin{align}
    d &= e^{-1} \mod \phi(n) \tag{2}
    \end{align}
    \item $\phi(n)$ is Euler's totient function, as shown in Equation (2).
    \item $n$ is the product of two prime numbers $p$ and $q$, used to generate the key.
\end{itemize}

This decryption process works based on the fact that:

\begin{align}
m &= (m^e)^d \mod n \tag{3}
\end{align}

Because $e \times d = 1 \mod \phi(n)$, it follows that:

\begin{align}
m &= m^{(e \times d)} \mod n = m \mod n \tag{4}
\end{align}

This returns the original value of $m$, allowing the AES key to be restored to its original form.

The process then continues with AES decryption, which uses the function \texttt{EVP\_DecryptUpdate()}. This function involves three stages, similar to the encryption process: the initial round, the main round, and the final round.

\begin{algorithm}
\caption{RSA Decryption Process}
\begin{algorithmic}[1]
\State \textbf{Input:} Ciphertext $c$, Private Key $(d, n)$
\State \textbf{Output:} Plaintext $m$ (retrieved AES key)

\State \textbf{Step 1: RSA Decryption}
\State Decrypt the ciphertext $c$ using the formula:
\[
m = c^d \mod n
\]

\State \textbf{Step 2: Relation Between Exponents}
\State This decryption is based on the relationship:
\[
m = (m^e)^d \mod n
\]

\State \textbf{Step 3: Simplification}
\State Since $e \times d = 1 \mod \phi(n)$, the result simplifies to:
\[
m = m \mod n
\]

\State \textbf{Output:} The plaintext $m$ is the original AES key.
\end{algorithmic}
\end{algorithm}

\begin{itemize}
\item \textbf{Initial Round}\\
In the initial round, the process starts with the AddRoundKey step before the first main round begins. In this step, the ciphertext is XORed with the round key, which can be expressed mathematically as shown in Equation (1):
\begin{align}
C_n &= C_{n-1} \oplus \text{Key}_{10} \tag{1}
\end{align}

Where:
\begin{itemize}
    \item $C_n$ represents the ciphertext after the initial round, as described in Equation (1).
    \item $C_{n-1}$ represents the original ciphertext, also used in Equation (1).
    \item $\text{Key}_{10}$ is the round key for the 10th round, as shown in Equation (1).
\end{itemize}

The AddRoundKey step ensures that the ciphertext $C_{n-1}$ is XORed with the round key $\text{Key}_{10}$, resulting in the new ciphertext $C_n$ that will be used in the following round of decryption.
\end{itemize}

\begin{itemize}
    \item \textbf{Main Round}\\
In this step, four functions must be applied: Inverse ShiftRows, Inverse SubBytes, AddRoundKey, and Inverse MixColumns. The Inverse ShiftRows function is the reverse of the ShiftRows operation, which conceptually does not involve a mathematical formula but rather a byte-shifting operation. For example, in the case of a 4x4 state matrix:

\begin{itemize}
    \item Row 1: No shift (remains the same).
    \item Row 2: Shifted 1 byte to the right.
    \item Row 3: Shifted 2 bytes to the right.
    \item Row 4: Shifted 3 bytes to the right.
\end{itemize}

This process can be expressed by the following formula, as shown in Equation (1):
\begin{align}
S_{\text{out}} &= \text{InvShiftRows}(S_{\text{in}}) \tag{1}
\end{align}

Where:
\begin{itemize}
    \item $S_{\text{out}}$ represents the state matrix after the inverse shift, as shown in Equation (1).
    \item $S_{\text{in}}$ represents the state matrix before the inverse shift.
\end{itemize}

Next, the AddRoundKey function is applied using the same formula as during encryption. After that, the Inverse SubBytes function is applied using the inverse S-Box. The formula for Inverse SubBytes is shown in Equation (2):

\begin{align}
S_{\text{out}}[i,j] &= S^{-1}(S_{\text{in}}[i,j]) \tag{2}
\end{align}

Where:
\begin{itemize}
    \item $S_{\text{in}}[i,j]$ is the byte from the state matrix at position $(i,j)$ before substitution, as shown in Equation (2).
    \item $S_{\text{out}}[i,j]$ is the substituted byte at the same position, as shown in Equation (2).
\end{itemize}

This step reverses the SubBytes operation performed during encryption, transforming the substituted values back to their original state using the inverse S-Box.

The last step is the Inverse MixColumns function, which can be expressed by the formula in Equation (3):

\begin{align}
c' &= M^{-1} \times c \tag{3}
\end{align}

Where:
\begin{itemize}
    \item $M^{-1}$ is the inverse mix columns matrix, as shown in Equation (3).
    \item $c$ represents a column from the state matrix before the inverse mix columns operation.
    \item $c'$ represents the column after the inverse mix columns operation, as shown in Equation (3).
\end{itemize}
\end{itemize}

\begin{itemize}
    \item \textbf{Final Rounds} \\
    In the final round, several functions are combined to complete the decryption process. It begins with Inverse ShiftRows, followed by Inverse SubBytes, and concludes with AddRoundKey. As a result, the data is returned to its original plaintext form, restoring it to its state before the encryption process begins. This final round ensures the data is fully decrypted and identical to the original input.
\end{itemize}

\begin{algorithm}
\caption{AES Decryption Process}
\begin{algorithmic}[1]
\State \textbf{Input:} Ciphertext $C$, Round Key $\text{Key}_{10}$
\State \textbf{Output:} Plaintext

\State \textbf{Initial Round:}
\State XOR the original ciphertext $C_{n-1}$ with the round key $\text{Key}_{10}$:
\[
C_n = C_{n-1} \oplus \text{Key}_{10}
\]

\State \textbf{Main Round:}
\State Perform the following steps for each round:
\begin{itemize}
    \item Apply Inverse ShiftRows:
    \[
    S_{\text{out}} = \text{InvShiftRows}(S_{\text{in}})
    \]
    
    \item Apply AddRoundKey (same formula as in encryption).
    
    \item Apply Inverse SubBytes:
    \[
    S_{\text{out}}[i,j] = S^{-1}(S_{\text{in}}[i,j])
    \]
    
    \item Apply Inverse MixColumns:
    \[
    c' = M^{-1} \times c
    \]
\end{itemize}

\State \textbf{Final Rounds:}
\State Combine the following steps to complete the decryption:
\begin{itemize}
    \item Apply Inverse ShiftRows
    \item Apply Inverse SubBytes
    \item Apply AddRoundKey
\end{itemize}

\State \textbf{Output:} The data is now restored to its original plaintext form.
\end{algorithmic}
\end{algorithm}

\subsection{System Evaluation}
\begin{figure} [h] 
    \centering
    \includegraphics[width = 7.5 cm]{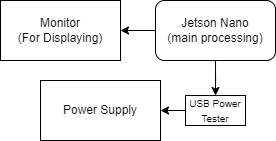}
 \caption{HSM Implementation}
    \label{fig/implementation}
\end{figure}

The testing of the system focuses on the performance of the Jetson Nano single-board computer, as shown in Fig. \ref{fig/implementation}. Several variables are used to assess the performance of the Jetson Nano, including the time consumption required for the data security process, such as encryption, the process of adding slots to SoftHSM, and decryption. Other variables include the memory load required when running the program, the power consumption needed during processing, and the final test, which ensures the system's security. This includes an attack test for key extraction, where attempts are made to extract the key from the SoftHSM slot forcibly \cite{li_defend_2024}.

These testing variables are selected based on the research conducted by Yilmaz \& Ozdemir (2018) \cite{yilmaz_performance_2018}. In their research, Yilmaz stated that one of the key factors in a hardware security module is the power consumption required to perform a process. Additionally, Jaspin et al. (2021), in their research, also conducted tests on time consumption during the processing of a hardware security module \cite{jaspin_efficient_2021}.

\section{Result} \label{results}
The results of several HSM tests are as follows:
\subsection{Time Consumption}
Time consumption testing is one of the key variables in processing a hardware security module (HSM). The time consumption of an HSM must be capable of processing data quickly \cite{yilmaz_performance_2018}. This test was conducted using the Htop tool on the Linux operating system running on a Jetson Nano. The calculation of time consumption involved measuring several types of time, including real-time, user time, and system time. Real-time refers to the actual time required to process the main data. User time is the CPU time used for data processing also known as user time. System time represents the amount of CPU time the operating system's kernel uses during data processing. This test was conducted with ten batches of data, each with 8 sample trials measured as a package (real-time, user time, and system time).

\begin{table}
\centering
\caption{Result time consumption for Encryption}
\begin{tabular}{|c|c|c|c|}
\hline
\textbf{Batch} & \textbf{Real (s)} & \textbf{User (s)} & \textbf{System (s)} \\
\hline
1  & 4.99  & 0.39 & 0.47 \\
2  & 2.95  & 0.39 & 0.47 \\
3  & 3.01  & 0.39 & 0.48 \\
4  & 2.81  & 0.40 & 0.47 \\
5  & 3.04  & 0.40 & 0.47 \\
6  & 3.30  & 0.41 & 0.48 \\
7  & 3.58  & 0.40 & 0.48 \\
8  & 2.91  & 0.38 & 0.47 \\
9  & 2.92  & 0.40 & 0.46 \\
10 & 3.41  & 0.41 & 0.46 \\
\hline
\textbf{Average} & \textbf{3.293} & \textbf{0.397} & \textbf{0.472} \\
\hline
\end{tabular}
\end{table}

The results of the time consumption test show that the double encryption process for a 500 MB file requires an average of 3.293 seconds in real time, 0.397 seconds in user time, and 0.472 seconds in system time.

These results indicate that encryption processes like AES and RSA, which involve loops of randomization and permutation, require complex calculations. Processing time will naturally increase as the number of files grows. Additionally, system time will increase as the operating system manages more memory for the processed files. Similarly, user time will also rise as the CPU handles more operations.

\begin{table}
\centering
\caption{Result time consumption for Decryption}
\begin{tabular}{|c|c|c|c|}
\hline
\textbf{Batch} & \textbf{Real (s)} & \textbf{User (s)} & \textbf{System (s)} \\
\hline
1  & 2.51  & 0.074 & 0.453 \\
2  & 2.595 & 0.073 & 0.451 \\
3  & 2.580 & 0.075 & 0.453 \\
4  & 2.514 & 0.079 & 0.450 \\
5  & 2.570 & 0.076 & 0.449 \\
6  & 2.587 & 0.074 & 0.449 \\
7  & 2.541 & 0.078 & 0.450 \\
8  & 2.590 & 0.074 & 0.448 \\
9  & 2.555 & 0.075 & 0.451 \\
10 & 2.587 & 0.074 & 0.449 \\
\hline
\textbf{Average} & \textbf{2.558} & \textbf{0.076} & \textbf{0.452} \\
\hline
\end{tabular}
\end{table}

The results of the time consumption test for the decryption process show that the average time required in real-time to process the decryption is 2.558 seconds, while the user time is 0.0762 seconds, and the system time is 0.452 seconds.

This average time is faster than the encryption process, which took 3.293 seconds in real time, 0.397 seconds in user time, and 0.472 seconds in system time.

These results are also higher by 0.5 seconds compared to the study presented by Jaspin et al. Detailed comparisons are available in Table 1 for encryption and Table 2 for decryption, which compares the conversion results of encryption and decryption between the algorithm developed in this research and the one previously developed by Jaspin et al (Table 3 \& Table 4). This shows that the algorithm developed in this study has good time efficiency compared to previous research.

\begin{table*}
\centering
\caption{Comparative Study of time consumption for encryption 
\cite{jaspin_efficient_2021}}
\begin{tabular}{|c|c|c|c|c|c|c|}
\hline
\textbf{File (MB)} & \textbf{DES (s)} & \textbf{Blowfish (s)} & \textbf{RC5 (s)} & \textbf{3DES (s)} & \textbf{AES + RSA \cite{jaspin_efficient_2021}} & \textbf{AES + RSA} \\
\hline
0.1  & 3  & 1.5  & 2  & 2  & 1  & 0.03 \\
0.3  & 4  & 2.1  & 2.5  & 2.4  & 1.5  & 0.05 \\
0.5  & 4.2  & 2.5  & 3  & 3.1  & 2.1  & 0.1 \\
0.75 & 4.5  & 3  & 3.5  & 3.6  & 3  & 0.4 \\
1    & 5  & 3.2  & 4  & 4.2  & 4  & 0.5 \\
500  & -  & -  & -  & 5  & -  & - \\
\hline
\end{tabular}
\end{table*}

\begin{table*}
\centering
\caption{Comparative Study of time consumption for decryption 
\cite{jaspin_efficient_2021}}
\begin{tabular}{|c|c|c|c|c|c|c|}
\hline
\textbf{File (MB)} & \textbf{DES (s)} & \textbf{Blowfish (s)} & \textbf{RC5 (s)} & \textbf{4-DES (s)} & \textbf{AES + RSA \cite{jaspin_efficient_2021}} & \textbf{AES + RSA} \\
\hline
0.1  & 2.0  & 1.2  & 1.8  & 1  & 0.9  & 0.03 \\
0.3  & 2.5  & 1.5  & 2.3  & 1.5  & 1.2  & 0.04 \\
0.5  & 3  & 1.7  & 2.7  & 2.1  & 1.5  & 0.1 \\
0.75 & 3.5  & 2  & 3  & 2.5  & 2.3  & 0.2 \\
1    & 4  & 2.2  & 3.5  & 3  & 2.9  & 0.4 \\
500  & -  & -  & -  & -  & -  & - \\
\hline
\end{tabular}
\end{table*}

\subsection{Memory Consumption}
Memory consumption is a crucial aspect of the use of HSM. In their research, Yilmaz et al. stated that the less memory load an HSM requires, the better its performance \cite{yilmaz_performance_2018}. In this study, memory consumption testing is divided into two categories: primary memory consumption and CPU memory consumption. Main memory, or RAM, stores data and instructions for programs running on the computer. Efficient memory management in this context is essential to ensure smooth and uninterrupted system performance, especially during high-demand tasks.
CPU memory stores copies of data, which are used to run programs that may be needed by the CPU in the near future. Both measurements used the Htop feature on the Ubuntu 20.04 operating system. This test calculated the main memory and CPU memory usage when processing a 500 MB file during encryption and decryption. The experiment was conducted by collecting data over ten batches, with each batch consisting of 8 sample trials measuring both memory and CPU memory consumption as a package (Table V).

\begin{table}[h!]
\centering
\caption{Result testing of memory consumption for encryption and decryption}
\begin{tabular}{|c|c|c|c|c|}
\hline
\multirow{2}{*}{\textbf{Batch}} & \multicolumn{2}{c|}{\hspace{10pt} \textbf{Encryption} \hspace{10pt}} & \multicolumn{2}{c|}{\hspace{10pt} \textbf{Decryption} \hspace{10pt}} \\ \cline{2-5} 
                       & \textbf{CPU (\%)}      & \textbf{Memory (\%)}    & \textbf{CPU (\%)}      & \textbf{Memory (\%)}    \\ \hline
1                      & 20.042        & 0.1            & 22.287        & 4.975          \\ \hline
2                      & 33.025        & 0.1            & 27.012        & 3.000          \\ \hline
3                      & 43.825        & 0.1            & 25.162        & 1.987          \\ \hline
4                      & 42.487        & 0.1            & 22.127        & 1.337          \\ \hline
5                      & 37.112        & 0.1            & 22.837        & 2.000          \\ \hline
6                      & 40.200        & 0.1            & 26.762        & 1.725          \\ \hline
7                      & 43.987        & 0.1            & 24.265        & 1.725          \\ \hline
8                      & 31.650        & 0.1            & 24.275        & 2.112          \\ \hline
9                      & 40.662        & 0.1            & 25.062        & 2.375          \\ \hline
10                     & 39.450        & 0.1            & 23.062        & 2.175          \\ \hline
\textbf{Average}        & \textbf{37.240}  & \textbf{0.1} & \textbf{24.243}  & \textbf{2.535} \\ \hline
\end{tabular}
\end{table}

During the encryption process, the memory consumption test results show that the average primary memory consumption from processing ten batches of data is 37.24\% of the total main memory.

Meanwhile, CPU memory remained stable during data processing, at 0.1\%. This outcome is attributed to two factors: efficient memory usage and efficient temporary storage. Efficient memory usage refers to how the operating system dynamically allocates and frees memory to optimize resource utilization.

Efficient temporary storage or CPU system efficiency means that the cryptographic program uses RAM for temporary data storage \cite{wei_time_2024} \cite{liu_atvitsc_2024}. Once the process is complete, the memory of this data is cleared. Table 5 shows the memory consumption test results for the decryption process.

The testing shows that the average primary memory consumption during the decryption process is 24.24\%, lower than the encryption process, which reached 37.24\%. However, the CPU memory consumption during the decryption process is 2.535\%, significantly higher than the encryption process, which remained at 0.1

This difference occurs because decryption involves restoring a large amount of processed data, requiring more CPU operations. Additionally, decryption includes more operations to reverse the encryption through complex inverse algorithms and state recovery.

In contrast, the encryption process uses more memory to store temporary states, while decryption demands more resources to restore these states optimally. These results indicate that the decryption process is more CPU-efficient and requires more memory resources than encryption.

\subsection{Power Consumption}
In this test, voltage and current changes on the Jetson Nano were observed during various processes, including encryption, idle state, and running SoftHSM. The test was conducted using a USB Voltage Current Meter Tester Power Detector connected to the Jetson Nano's power adapter cable (Table 6).

\begin{table}
\centering
\caption{Results of Power Consumption Testing on Jetson Nano}
\vspace{0.2 cm}
\begin{tabular}{|l|c|c|}
\hline
\textbf{Process}                     & \textbf{Voltage} & \textbf{Current} \\
\hline
Idle                                 & 5.19 V           & 0.41 A           \\
Running Text Code                    & 5.18 V           & 0.42 A           \\
Generate AES Encryption Code         & 5.22 V           & 0.87 A           \\
Running AES Encryption               & 5.30 V           & 1.21 A           \\
Generate RSA Key                     & 5.20 V           & 0.50 A           \\
Running RSA Encryption               & 5.22 V           & 0.58 A           \\
Connect to HSM Slot                  & 5.19 V           & 0.48 A           \\
Running RSA Decryption               & 5.20 V           & 0.66 A           \\
Generate AES Decryption Code         & 5.18 V           & 1.20 A           \\
Running AES Decryption               & 5.21 V           & 1.20 A           \\
\hline
\end{tabular}
\end{table}

The test results show that power consumption increases according to the tasks performed. In idle mode, power consumption was approximately 5.19 V with a current of 0.41 A. When running simple code, the current increased to 0.42 A. When generating AES keys for encryption, power consumption reached 5.22 V and 0.87 A. Full AES encryption required 5.30 V and 1.21 A, indicating a high computational load.

Generating RSA keys required less power, with 5.20 V and 0.50 A, while RSA encryption consumed 5.22 V and 0.58 A. When connecting to the HSM slot, power usage remained relatively low at 5.19 V and 0.48 A. RSA decryption slightly increased power consumption to 5.20 V and 0.66 A while generating AES keys for decryption reached 5.18 V and 1.20 A.

Overall, AES cryptographic operations required more power, while RSA and communication tasks were more efficient. AES processes large amounts of data, whereas RSA only encrypts the AES keys.

\subsection{Attacking Simulation}
\begin{figure} [h] 
    \centering
    \includegraphics[width = 9 cm]{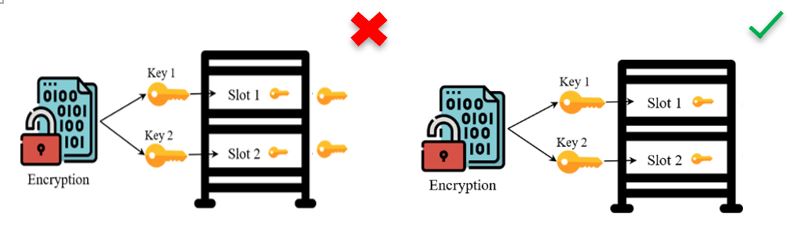}
 \caption{Attacking Concept}
    \label{fig/attack}
\end{figure}

SoftHSM has a solid security layer; once a private key is stored in a SoftHSM slot, it cannot be extracted. This is because an HSM must effectively safeguard the protected data. In this research, the researchers conducted a test to determine whether the slot system in SoftHSM provides a high level of security for protecting the RSA private key within the slot. The test involved a key extraction attack, simulating a man-in-the-middle scenario where an attempt is made to extract the RSA private key from the SoftHSM slot. The SoftHSM slot system can be considered insecure if the key can be extracted. However, if the key remains securely stored, the system can be deemed to have strong security. The illustration of this test is shown in Fig. \ref{fig/attack}

\begin{figure} [h] 
    \centering
    \includegraphics[width = 8.5 cm]{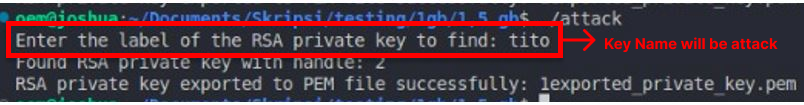}
 \caption{Extracting Key Process}
    \label{fig/extracting}
\end{figure}

The test used a key extraction program adapted to the PKCS11 interface and OpenSSL library. Initially, the program connects directly to the targeted slot and then prompts the user to input the correct key label to proceed. The program then attempts to extract the key as a PEM file, automatically naming it `1exported\_private\_key.pem`, as shown in Fig. \ref{fig/extracting}.

\begin{figure} [h] 
    \centering
    \includegraphics[width = 8.7 cm]{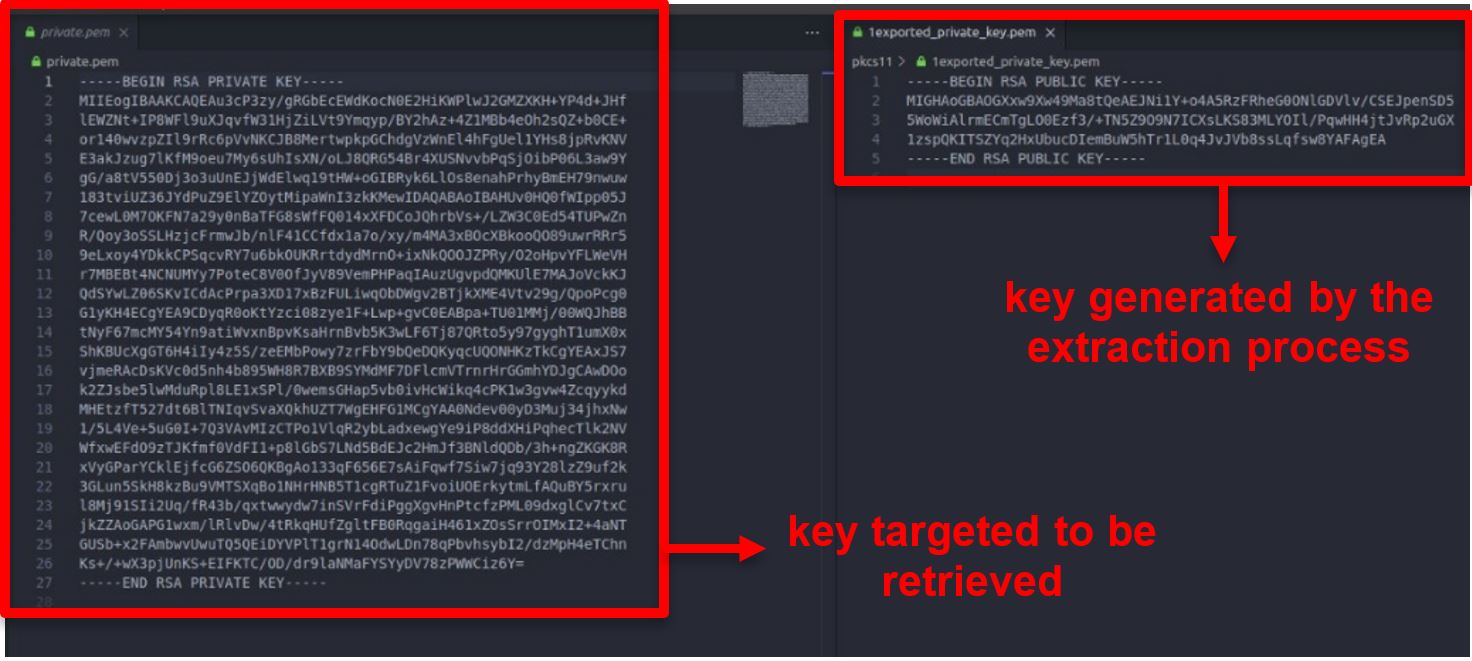}
 \caption{Attacking Result}
    \label{fig/result}
\end{figure}

During the test, while the extraction process occurred, as seen in Fig. \ref{fig/result}, the extracted key was not the intended private key but a random public key. The RSA private key remained securely stored in the slot without any damage. This indicates the key under attack was still safely stored within the SoftHSM slot. The protection system implemented for the SoftHSM slot functions similarly to a Honeypot, or as described in the research by Naik and Jenkins, where the system deceives attackers, ensuring that the data remains secure. While the system may appear to extract a key, it does not release the private key, thus keeping the data safe within the slot \cite{saed_detection_2022}.

\subsection{Data Processing Results}

The most critical component in data utilization is maintaining the integrity and authenticity of the data \cite{han_privacy-preserving_2024} \cite{miao_blockchain-based_2024}. This forms the basis for the current test being conducted. This test compares data before and after being processed inside the HSM. The sample data being compared is in the form of image data. Image data is chosen because it makes it easier to see if there is a visual change in the data after processing. The comparison is based on visual aspects and the data size before and after processing. The results of this research are shown in Table 7.

\begin{table} 
\centering
\caption{Data Processing System Results}
\vspace{0.2 cm}
\begin{tabular}{|c|c|}
\hline
\textbf{Before Processing} & \textbf{After Processing} \\
\hline
\includegraphics[width=0.2\textwidth]{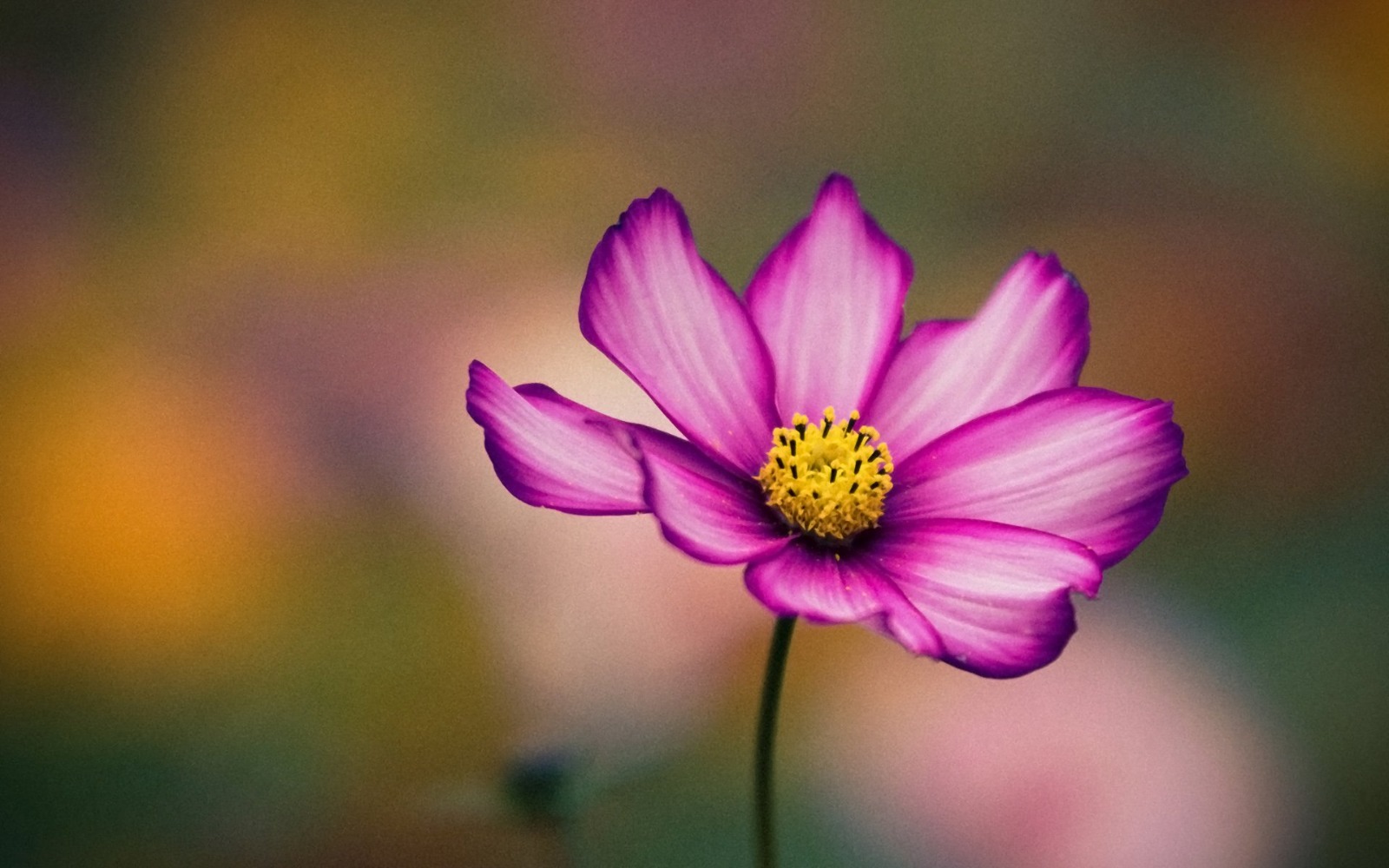} & \includegraphics[width=0.2\textwidth]{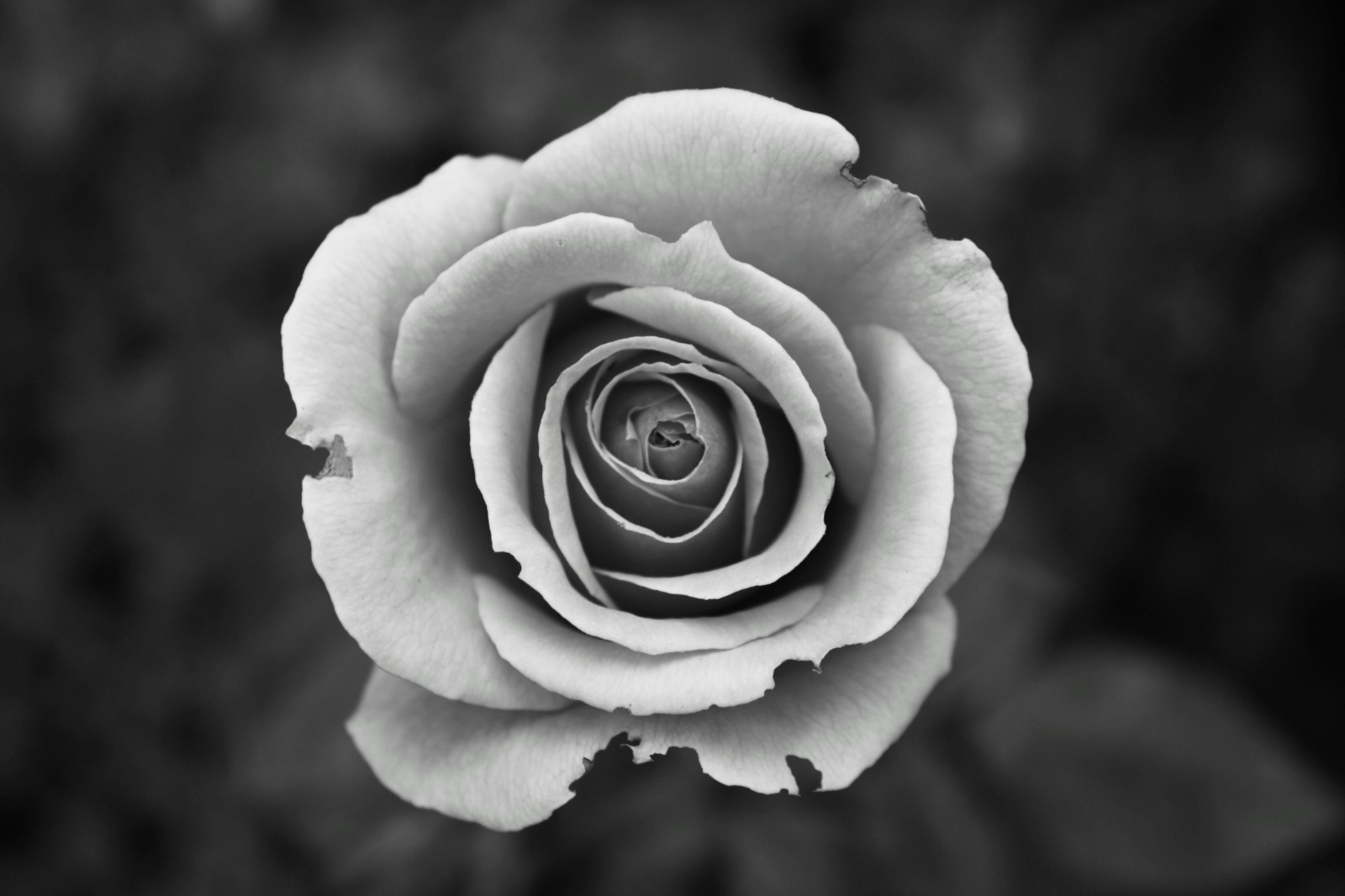} \\
\hline
\includegraphics[width=0.2\textwidth]{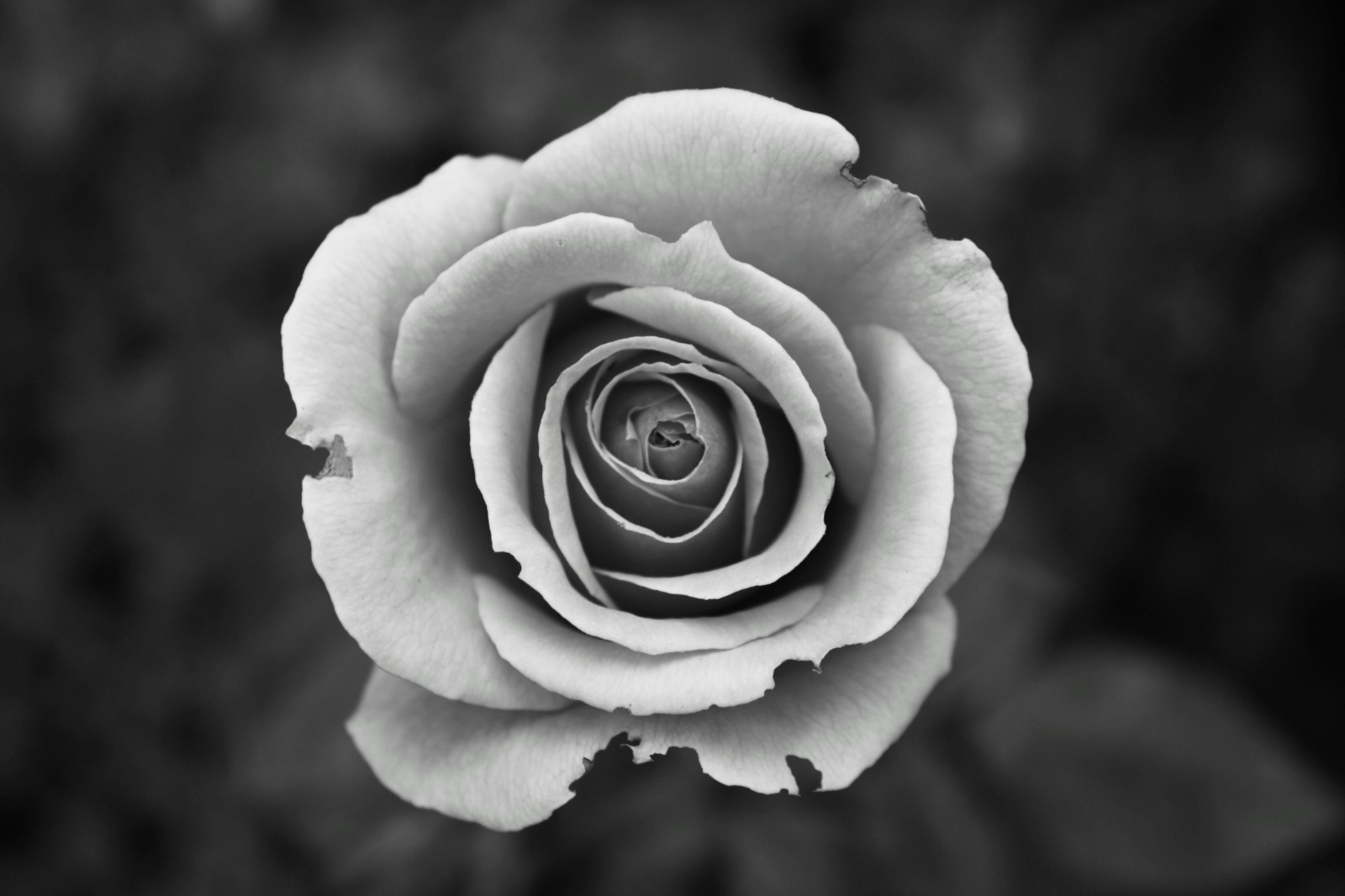} & \includegraphics[width=0.2\textwidth]{figs/bunga_terenkripsi_1.jpg} \\
\hline
\includegraphics[width=0.2\textwidth]{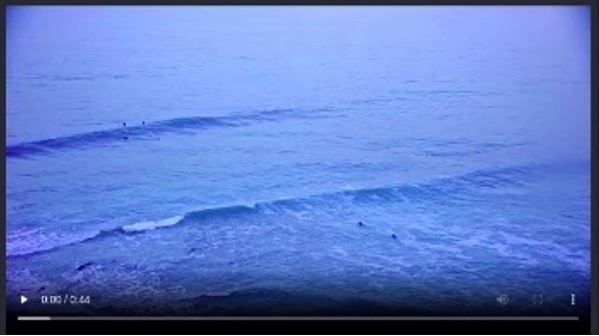} & \includegraphics[width=0.2\textwidth]{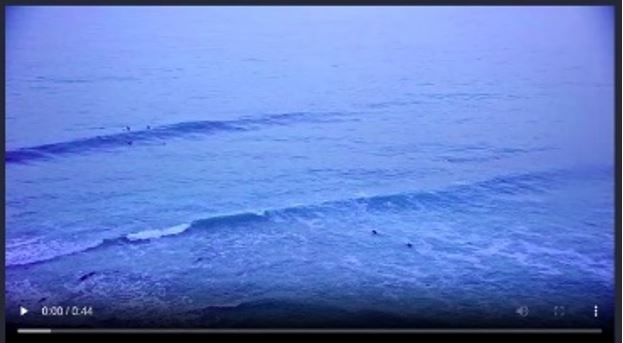} \\
\hline
\end{tabular}
\end{table}

The system successfully processed colored video and was able to use a 4096-bit key size. The results of this test show that the data processed before and after being handled by the HSM did not undergo any changes in terms of visualization or data size. This is confirmed by Table VIII, which shows no visual changes to the data. However, during the data decryption process, it is essential to consider the key size being generated; if it is too small, the data may not be decrypted correctly. On the other hand, the size of the processed data, such as 500 MB, did not change after being processed in the HSM.

Several factors influence the absence of data corruption: AES uses symmetric keys, and RSA employs public-private keys. This structured process ensures that the correct key is used so the data will always return to its original form. Additionally, AES and RSA operate at the bit level during complex transformations to ensure security, spreading the influence of each bit across the entire cipher block, thus reducing the risk of data corruption. The results of this test prove that the data processed within the HSM can be guaranteed to remain intact with no damage to the processed data. 

\section{Conclusions} \label{conclusions}
The HSM system successfully protected against man-in-the-middle attacks, such as key extraction, ensuring the RSA private key remained secure. The HSM could process data without changing visibility, size, or overall data integrity. The system also demonstrated strong performance, with an average encryption time of 3.29 seconds, slot access time of 0.018 seconds, and decryption time of 2.558 seconds. Memory usage was efficient, with 37.24\% encryption and 24.24\% decryption. The average power consumption of the HSM during all processes was 5.20 V and 0.72 A.

\ifCLASSOPTIONcompsoc
  \section*{Acknowledgments}
\else
  \section*{Acknowledgment}
\fi

This work was partially supported by the Department of Computer Science and Electronics, Universitas Gadjah Mada under the publication Funding Year 2024.

\ifCLASSOPTIONcaptionsoff
  \newpage
\fi




%
\bibliographystyle{IEEEtran}
\bibliography{citation}

\begin{thebibliography}{10}
\providecommand{\url}[1]{#1}
\csname url@samestyle\endcsname
\providecommand{\newblock}{\relax}
\providecommand{\bibinfo}[2]{#2}
\providecommand{\BIBentrySTDinterwordspacing}{\spaceskip=0pt\relax}
\providecommand{\BIBentryALTinterwordstretchfactor}{4}
\providecommand{\BIBentryALTinterwordspacing}{\spaceskip=\fontdimen2\font plus
\BIBentryALTinterwordstretchfactor\fontdimen3\font minus
  \fontdimen4\font\relax}
\providecommand{\BIBforeignlanguage}[2]{{%
\expandafter\ifx\csname l@#1\endcsname\relax
\typeout{** WARNING: IEEEtran.bst: No hyphenation pattern has been}%
\typeout{** loaded for the language `#1'. Using the pattern for}%
\typeout{** the default language instead.}%
\else
\language=\csname l@#1\endcsname
\fi
#2}}
\providecommand{\BIBdecl}{\relax}
\BIBdecl

\bibitem{frank_industry_2019}
A.~G. Frank, L.~S. Dalenogare, and N.~F. Ayala, ``Industry 4.0 technologies:
  {Implementation} patterns in manufacturing companies,'' \emph{Inter. J.
  Production Economics}, vol. 210, pp. 15--26, apr 2019.

\bibitem{dalenogare_expected_2018}
L.~S. Dalenogare and Benitez, ``The expected contribution of {Industry} 4.0
  technologies for industrial performance,'' \emph{International Journal of
  Production Economics}, vol. 204, pp. 383--394, oct 2018.

\bibitem{regla_performance_2022}
A.~I. Regla and E.~D. Festijo, ``Performance {Analysis} of {Light}-weight
  {Cryptographic} {Algorithms} for {Internet} of {Things} ({IoT})
  {Applications}: {A} {Systematic} {Review},'' in \emph{2022 {IEEE} 7th
  {International} conference for {Convergence} in {Technology} ({I2CT})}, apr
  2022, pp. 1--5.

\bibitem{dinlersoz_automation_2023}
E.~Dinlersoz and Z.~Wolf, ``\BIBforeignlanguage{en}{Automation, labor share,
  and productivity: plant-level evidence from {U}.{S}. manufacturing},''
  \emph{\BIBforeignlanguage{en}{Economics of Innovation and New Technology}},
  pp. 1--23, Jul. 2023.

\bibitem{yuan_leakage_2024}
B.~Yuan, M.~Yang, Z.~Xu, Q.~Chen, Z.~Song, Z.~Li, D.~Zou, and H.~Jin,
  ``\BIBforeignlanguage{en}{Leakage of {Authorization}-{Data} in {IoT} {Device}
  {Sharing}: {New} {Attacks} and {Countermeasure}},''
  \emph{\BIBforeignlanguage{en}{IEEE Trans. on Dependable and Secure
  Computing}}, vol.~21, no.~4, pp. 3196--3210, Jul. 2024.

\bibitem{schuh_industrie_nodate}
G.~Schuh, R.~Anderl, and Dumitrescu, ``Industrie 4.0 maturity index managing
  the digital transformation of companies,'' \emph{acatech-National Academy of
  Science and Engineering}, pp. 1--54, June 2020.

\bibitem{amael_securing_2024}
J.~T. Amael, J.~E. Istiyanto, Z.~Hakim, R.~H. Sari, A.~Z.~K. Frisky, and
  O.~Natan, ``Securing {Ventilators}: {Integrating} {Hardware} {Security}
  {Modules} with {SoftHSM} and {Cryptographic} {Algorithms},'' in \emph{2024
  {IEEE} 33rd {International} {Symposium} on {Industrial} {Electronics}
  ({ISIE})}, jun 2024, pp. 1--6.

\bibitem{r1}
A.~Brandao and P.~Georgieva, ``Log {Files} {Analysis} {For} {Network}
  {Intrusion} {Detection},'' in \emph{{IEEE} Inter. Conf. {Intelligent}
  {Systems} ({IS})}, aug 2020, pp. 328--333.

\bibitem{wang_file_2011}
S.~Wang and G.~Liu, ``\BIBforeignlanguage{en}{File {Encryption} and
  {Decryption} {System} {Based} on {RSA} {Algorithm}},'' in
  \emph{\BIBforeignlanguage{en}{Inter. Conf. {Computational} and {Information}
  {Sciences}}}.\hskip 1em plus 0.5em minus 0.4em\relax Chengdu, China: IEEE,
  Oct. 2011, pp. 797--800.

\bibitem{mallik_man---middle-attack_2019}
A.~Mallik, A.~Ahsan, M.~M.~Z. Shahadat, and J.-C. Tsou,
  ``\BIBforeignlanguage{en}{Man-in-the-middle-attack: {Understanding} in simple
  words},'' \emph{\BIBforeignlanguage{en}{Inter. J. Data and Network Science}},
  pp. 77--92, 2019.

\bibitem{saed_detection_2022}
M.~Saed and A.~Aljuhani, ``\BIBforeignlanguage{en}{Detection of {Man} in {The}
  {Middle} {Attack} using {Machine} learning},'' in
  \emph{\BIBforeignlanguage{en}{int. {Conference} on {Computing} and
  {Information} {Technology} ({ICCIT})}}.\hskip 1em plus 0.5em minus
  0.4em\relax Tabuk, Saudi Arabia: IEEE, Jan. 2022, pp. 388--393.

\bibitem{hupp_module-ot_2020}
W.~Hupp, A.~Hasandka, R.~S. De~Carvalho, and D.~Saleem,
  ``\BIBforeignlanguage{en}{Module-{OT}: {A} {Hardware} {Security} {Module} for
  {Operational} {Technology}},'' in \emph{\BIBforeignlanguage{en}{{IEEE} Inter.
  conf ({TPEC})}}.\hskip 1em plus 0.5em minus 0.4em\relax College Station, TX,
  USA: IEEE, Feb. 2020, pp. 1--6.

\bibitem{bathalapalli_itpm_2023}
V.~K. V.~V. Bathalapalli, S.~P. Mohanty, E.~Kougianos, V.~Iyer, and B.~Rout,
  ``\BIBforeignlanguage{en}{{iTPM}: {Exploring} {PUF}-based {Keyless} {TPM} for
  {Security}-by-{Design} of {Smart} {Electronics}},'' in
  \emph{\BIBforeignlanguage{en}{{IEEE} {Computer} {Society} {Annual} Symp.
  {VLSI} ({ISVLSI})}}.\hskip 1em plus 0.5em minus 0.4em\relax Foz do Iguacu,
  Brazil: IEEE, Jun. 2023, pp. 1--6.

\bibitem{benhani_security_2019}
E.~M. Benhani, L.~Bossuet, and A.~Aubert, ``\BIBforeignlanguage{en}{The
  {Security} of {ARM} {TrustZone} in a {FPGA}-{Based} {SoC}},''
  \emph{\BIBforeignlanguage{en}{IEEE Trans. Computers}}, vol.~68, no.~8, pp.
  1238--1248, Aug. 2019.

\bibitem{mulder_trends_2023}
V.~Mulder, A.~Mermoud, and Lenders, Eds., \emph{\BIBforeignlanguage{en}{Trends
  in {Data} {Protection} and {Encryption} {Technologies}}}.\hskip 1em plus
  0.5em minus 0.4em\relax Cham: Springer Nature Switzerland, 2023.

\bibitem{balan_puf-based_2020}
A.~Balan, T.~Balan, M.~Cirstea, and F.~Sandu, ``\BIBforeignlanguage{en}{A
  {PUF}-based cryptographic security solution for {IoT} systems on chip},''
  \emph{\BIBforeignlanguage{en}{EURASIP Journal on Wireless Communications and
  Networking}}, vol. 2020, no.~1, p. 231, Dec. 2020.

\bibitem{jaspin_efficient_2021}
K.~Jaspin, S.~Selvan, S.~Sahana, and G.~Thanmai,
  ``\BIBforeignlanguage{en}{Efficient and {Secure} {File} {Transfer} in {Cloud}
  {Through} {Double} {Encryption} {Using} {AES} and {RSA} {Algorithm}},'' in
  \emph{\BIBforeignlanguage{en}{Inter. Conf ({ESCI})}}.\hskip 1em plus 0.5em
  minus 0.4em\relax Pune, India: IEEE, Mar. 2021, pp. 791--796.

\bibitem{prawira_p_secure_2020}
M.~M. Prawira~P, R.~Kurniandi, and A.~Amiruddin,
  ``\BIBforeignlanguage{en}{Secure {SMS} {Using} {Pseudo}-{Random} {Bit}
  {Generator} {Based} on {Chaotic} {Map}, and {AES} on {Arduino} {UNO} {Board}
  and {SIM} 900 {Module}},'' in \emph{\BIBforeignlanguage{en}{Inter. Work
  ({IWBIS})}}.\hskip 1em plus 0.5em minus 0.4em\relax Depok, Indonesia: IEEE,
  Oct. 2020, pp. 91--96.

\bibitem{yilmaz_performance_2018}
B.~Yilmaz and S.~OzdemIr, ``\BIBforeignlanguage{tr}{Performance comparison of
  cryptographic algorithms in internet of things},'' in
  \emph{\BIBforeignlanguage{tr}{Inter. Conf ({SIU})}}.\hskip 1em plus 0.5em
  minus 0.4em\relax Izmir: IEEE, May 2018, pp. 1--4.

\bibitem{homma_toward_2014}
N.~Homma, K.~Saito, and T.~Aoki, ``\BIBforeignlanguage{en}{Toward {Formal}
  {Design} of {Practical} {Cryptographic} {Hardware} {Based} on {Galois}
  {Field} {Arithmetic}},'' \emph{\BIBforeignlanguage{en}{IEEE Trans.
  Computers}}, vol.~63, no.~10, pp. 2604--2613, Oct. 2014.

\bibitem{mulder_trends_2023-1}
V.~Mulder, A.~Mermoud, V.~Lenders, and B.~Tellenbach, Eds.,
  \emph{\BIBforeignlanguage{en}{Trends in {Data} {Protection} and {Encryption}
  {Technologies}}}.\hskip 1em plus 0.5em minus 0.4em\relax Cham: Springer
  Nature Switzerland, 2023.

\bibitem{rady_memristor-based_2019}
H.~Rady, H.~Hossam, M.~Saied, and H.~Mostafa,
  ``\BIBforeignlanguage{en}{Memristor-{Based} {AES} {Key} {Generation} for
  {Low} {Power} {IoT} {Hardware} {Security} {Modules}},'' in
  \emph{\BIBforeignlanguage{en}{{IEEE} Inter. {Midwest} Symp. {Circuits} and
  {Systems} ({MWSCAS})}}.\hskip 1em plus 0.5em minus 0.4em\relax Dallas, TX,
  USA: IEEE, Aug. 2019, pp. 231--234.

\bibitem{jafarzadeh_real_2020}
H.~Jafarzadeh and A.~Jahanian, ``\BIBforeignlanguage{en}{Real {Vulnerabilities}
  in {Partial} {Reconfigurable} {Design} {Cycles}; {Case} {Study} for
  {Implementation} of {Hardware} {Security} {Modules}},'' in
  \emph{\BIBforeignlanguage{en}{Inter. Symp. {Computer} {Architecture} and
  {Digital} {Systems} ({CADS})}}.\hskip 1em plus 0.5em minus 0.4em\relax Rasht,
  Iran: IEEE, Aug. 2020, pp. 1--4.

\bibitem{hu_engineering_2023}
J.~Hu, W.~Wang, K.~Gaj, L.~Wang, and H.~Wang,
  ``\BIBforeignlanguage{en}{Engineering {Practical} {Rank}-{Code}-{Based}
  {Cryptographic} {Schemes} on {Embedded} {Hardware}. {A} {Case} {Study} on
  {ROLLO}},'' \emph{\BIBforeignlanguage{en}{IEEE Trans. Computers}}, pp. 1--17,
  2023.

\bibitem{pott_firmware_2021}
C.~Pott, P.~Jungklass, D.~J. Csejka, T.~Eisenbarth, and M.~Siebert,
  ``\BIBforeignlanguage{en}{Firmware {Security} {Module}: {A} {Framework} for
  {Trusted} {Computing} in {Automotive} {Multiprocessors}},''
  \emph{\BIBforeignlanguage{en}{Journal of Hardware and Systems Security}},
  vol.~5, no.~2, pp. 103--113, Jun. 2021.

\bibitem{kuzminykh_analysis_2020}
I.~Kuzminykh, M.~Yevdokymenko, and D.~Ageyev,
  ``\BIBforeignlanguage{en}{Analysis of {Encryption} {Key} {Management}
  {Systems}: {Strengths}, {Weaknesses}, {Opportunities}, {Threats}},'' in
  \emph{\BIBforeignlanguage{en}{{IEEE} Inter. Conf. ({PIC} {S}\&{T})}}.\hskip
  1em plus 0.5em minus 0.4em\relax Kharkiv, Ukraine: IEEE, Oct. 2020, pp.
  515--520.

\bibitem{luo_tz-kms_2018}
S.~Luo, Z.~Hua, and Y.~Xia, ``\BIBforeignlanguage{en}{{TZ}-{KMS}: {A} {Secure}
  {Key} {Management} {Service} for {Joint} {Cloud} {Computing} with {ARM}
  {TrustZone}},'' in \emph{\BIBforeignlanguage{en}{{IEEE} Symp.{System}
  {Engineering}}}.\hskip 1em plus 0.5em minus 0.4em\relax Bamberg: IEEE, Mar.
  2018, pp. 180--185.

\bibitem{kim_reinforcement_2022}
B.~Kim, S.~Yoon, and Y.~Kang, ``\BIBforeignlanguage{en}{Reinforcement of {IoT}
  {Open} {Platform} {Security} using {PUF} -based {Device} {Authentication}},''
  in \emph{\BIBforeignlanguage{en}{Inter. Conf ({ICTC})}}.\hskip 1em plus 0.5em
  minus 0.4em\relax Jeju Island, Korea, Republic of: IEEE, Oct. 2022, pp.
  1969--1971.

\bibitem{yoon_security_2023}
S.~Yoon, B.~Kim, and Y.~Kang, ``\BIBforeignlanguage{en}{Security enhancement
  scheme for mobile device using {H}/{W} cryptographic module},'' in
  \emph{\BIBforeignlanguage{en}{Inter. Conf ({ICTC})}}.\hskip 1em plus 0.5em
  minus 0.4em\relax Jeju Island, Korea, Republic of: IEEE, Oct. 2023, pp.
  1450--1452.

\bibitem{hassan_enhanced_2020}
M.~Hassan, A.~Sultan, A.~A. Awan, S.~Tahir, and I.~Ihsan,
  ``\BIBforeignlanguage{en}{An {Enhanced} and {Secure} {Multiserver}-based
  {User} {Authentication} {Protocol}},'' in \emph{\BIBforeignlanguage{en}{IEE
  Inter. Conf ({ICCWS})}}.\hskip 1em plus 0.5em minus 0.4em\relax Islamabad,
  Pakistan: IEEE, Oct. 2020, pp. 1--6.

\bibitem{yu_cryptanalysis_2023}
S.~Yu, K.~Kim, K.~Taesung, B.~Chung, and Y.~Kang,
  ``\BIBforeignlanguage{en}{Cryptanalysis and {Countermeasures} of the {Recent}
  {Authentication} and {Key} {Agreement} {Scheme} for {Internet} of
  {Drones}},'' in \emph{\BIBforeignlanguage{en}{2023 14th {International}
  {Conference} on {Information} and {Communication} {Technology} {Convergence}
  ({ICTC})}}.\hskip 1em plus 0.5em minus 0.4em\relax Jeju Island, Korea,
  Republic of: IEEE, Oct. 2023, pp. 1419--1422.

\bibitem{etibou_iot_2024}
J.~G.~V. Etibou and S.~Pierre, ``\BIBforeignlanguage{en}{{IoT} {Devices}
  {Modular} {Security} {Approach} {Using} {Positioning} {Security} {Engine}},''
  \emph{\BIBforeignlanguage{en}{IEEE Access}}, pp. 1--1, 2024.

\bibitem{luo_fast_2024}
W.~Luo, G.~Xie, Y.~Liu, X.~Xiao, and R.~Li, ``\BIBforeignlanguage{en}{Fast
  {Game} {Verification} for {Safety}- and {Security}-{Critical} {Distributed}
  {Applications}},'' \emph{\BIBforeignlanguage{en}{IEEE Trans. on Dependable
  and Secure Computing}}, pp. 1--18, 2024.

\bibitem{wang_building_2024}
C.~Wang, Y.~Deng, Z.~Ning, K.~Leach, J.~Li, S.~Yan, Z.~He, J.~Cao, and
  F.~Zhang, ``\BIBforeignlanguage{en}{Building a {Lightweight} {Trusted}
  {Execution} {Environment} for {Arm} {GPUs}},''
  \emph{\BIBforeignlanguage{en}{IEEE Trans. on Dependable and Secure
  Computing}}, pp. 1--16, 2024.

\bibitem{cai_secfed_2024}
Y.~Cai, W.~Ding, Y.~Xiao, Z.~Yan, X.~Liu, and Z.~Wan,
  ``\BIBforeignlanguage{en}{{SecFed}: {A} {Secure} and {Efficient} {Federated}
  {Learning} {Based} on {Multi}-{Key} {Homomorphic} {Encryption}},''
  \emph{\BIBforeignlanguage{en}{IEEE Trans. on Dependable and Secure
  Computing}}, vol.~21, no.~4, pp. 3817--3833, Jul. 2024.

\bibitem{li_defend_2024}
K.~Li, C.~Baird, and D.~Lin, ``\BIBforeignlanguage{en}{Defend {Data}
  {Poisoning} {Attacks} on {Voice} {Authentication}},''
  \emph{\BIBforeignlanguage{en}{IEEE Trans. on Dependable and Secure
  Computing}}, vol.~21, no.~4, pp. 1754--1769, Jul. 2024.

\bibitem{wei_time_2024}
C.~Wei, G.~Hong, A.~Wang, J.~Wang, S.~Sun, Y.~Ding, L.~Zhu, and W.~Ma,
  ``\BIBforeignlanguage{en}{Time is not enough: {Timing} {Leakage} {Analysis}
  on {Cryptographic} {Chips} via {Plaintext}-{Ciphertext} {Correlation} in
  {Non}-timing {Channel}},'' \emph{\BIBforeignlanguage{en}{IEEE Trans. on
  Information Forensics and Security}}, pp. 1--1, 2024.

\bibitem{liu_atvitsc_2024}
Y.~Liu, X.~Wang, B.~Qu, and F.~Zhao, ``\BIBforeignlanguage{en}{{ATVITSC}: {A}
  novel encrypted traffic classification method based on deep learning},''
  \emph{\BIBforeignlanguage{en}{IEEE Trans. on Information Forensics and
  Security}}, pp. 1--1, 2024.

\bibitem{han_privacy-preserving_2024}
J.~Han, L.~Chen, A.~Hu, L.~Chen, and J.~Li,
  ``\BIBforeignlanguage{en}{Privacy-{Preserving} {Decentralized} {Functional}
  {Encryption} for {Inner} {Product}},'' \emph{\BIBforeignlanguage{en}{IEEE
  Trans. on Dependable and Secure Computing}}, vol.~21, no.~4, pp. 1680--1694,
  Jul. 2024.

\bibitem{miao_blockchain-based_2024}
Y.~Miao, K.~Gai, L.~Zhu, K.-K.~R. Choo, and J.~Vaidya,
  ``\BIBforeignlanguage{en}{Blockchain-{Based} {Shared} {Data} {Integrity}
  {Auditing} and {Deduplication}},'' \emph{\BIBforeignlanguage{en}{IEEE Trans.
  on Dependable and Secure Computing}}, vol.~21, no.~4, pp. 3688--3703, Jul.
  2024.

\end{thebibliography}

%

\begin{IEEEbiography}[{\includegraphics[width=1in,height=1.25in,clip,keepaspectratio]{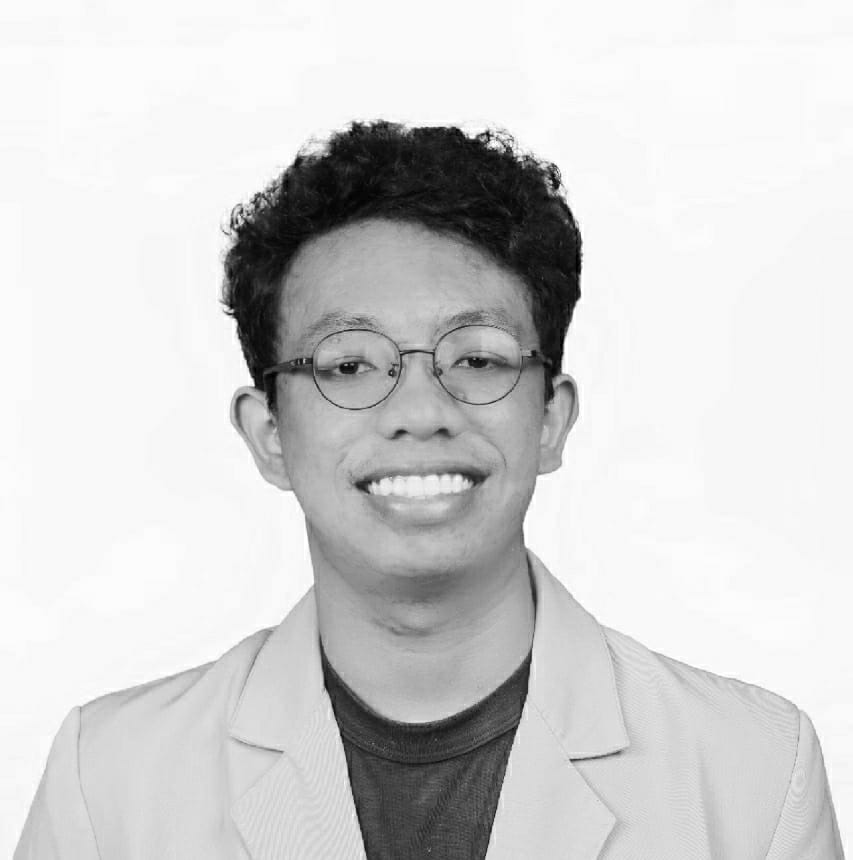}}]{Joshua Tito}
(Member, IEEE) is an undergraduate student at Electronic and Instrumentation, Universitas Gadjah Mada, Indonesia, in 2021. He has been actively involved in research and development in the fields of hardware programming, cybersecurity, and artificial intelligence. In 2024, he was recognized as an awardee best paper for students at the 33rd International Symposium on Industrial Electronics (ISIE) in Ulsan, South Korea. Since February 2024, he has served as an Assistant at the Electronics and Instrumentation Lab at Universitas Gadjah Mada. His research interests include hardware security modules, FPGA, and cryptographic systems.
\end{IEEEbiography}

\begin{IEEEbiography}[{\includegraphics[width=1in,height=1.25in,clip,keepaspectratio]{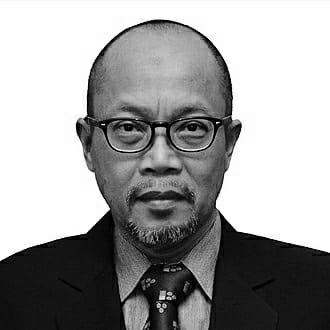}}]{Jazi Eko Istiyanto}
(Member, IEEE) received his B.Sc. in Physics from Universitas Gadjah Mada (UGM), Indonesia, in 1986. Then, he received his M.Sc. in Computer Science and Ph.D. in Electronic Systems Engineering from the University of Essex, UK, in 1988 and 1995, respectively. Since March 1988, he has been serving as a Lecturer at UGM. In August 2010, he became a Professor of Electronics and Instrumentation at the Department of Computer Science and Electronics, Faculty of Mathematics and Natural Sciences, UGM. He was appointed the 5th Head of the Republic of Indonesia's Nuclear Energy Regulatory Agency (BAPETEN) on February 7, 2014. To date, Prof. Jazi has authored and co-authored many papers published in internationally reputable journals and conferences. His research interests include embedded systems, cyber-security, and performance evaluation.

\end{IEEEbiography}

\begin{IEEEbiography}[{\includegraphics[width=1in,height=1.25in,clip,keepaspectratio]{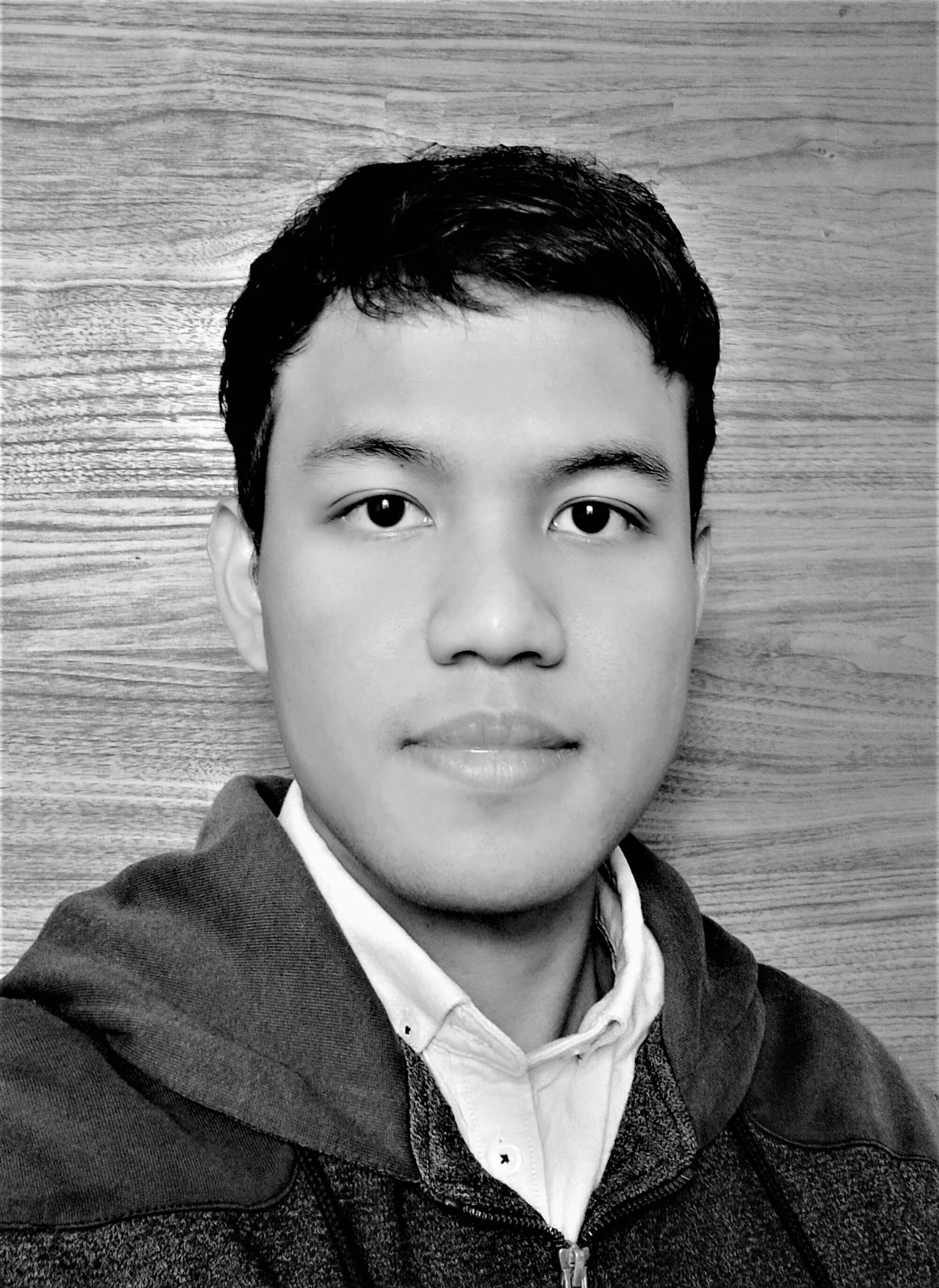}}]{Oskar Natan}
(Member, IEEE) received his B.A.Sc. in Electronics Engineering and M.Eng. in Electrical Engineering from Politeknik Elektronika Negeri Surabaya, Indonesia, in 2017 and 2019, respectively. In 2023, he received his Ph.D.(Eng.) in Computer Science and Engineering from Toyohashi University of Technology, Japan. Since January 2020, he has been affiliated with the Department of Computer Science and Electronics, Universitas Gadjah Mada, Indonesia, first as a Lecturer and currently serves as an Assistant Professor. His research interests lie in the fields of sensor fusion, hardware acceleration, and end-to-end systems for various computer science and electronics applications.
\end{IEEEbiography}
\vfill




\end{document}